\documentclass[journal,onecolumn,12pt,letterpaper]{IEEEtran}
\usepackage{graphicx,psfrag,subfigure,url,amsmath,amsfonts,
epsfig,amssymb}

\newtheorem{theorem}{Theorem} 
\newtheorem{proposition}{Proposition} 
\newtheorem{corollary}{Corollary}

\newcommand{\Prob}{{\rm P}}
\long\def\symbolfootnote[#1]#2{\begingroup
\def\thefootnote{\fnsymbol{footnote}}\footnote[#1]{#2}\endgroup}

\begin{document}
\title{The Capacity Region of a Class of 3-Receiver Broadcast Channels with
Degraded Message Sets}

\author{Chandra~Nair,~\IEEEmembership{Member,~IEEE,}
		and~Abbas~El~Gamal,~\IEEEmembership{Fellow,~IEEE}
\thanks{Chandra Nair was partly supported by the Direct Grant for research at
the Chinese University of Hong Kong}}

\maketitle

\begin{abstract}
K\"{o}rner and Marton established the capacity region for the 2-receiver
broadcast channel with degraded message sets. Recent results and conjectures 
suggest that a straightforward extension of the K\"{o}rner-Marton region to more 
than 2 receivers is optimal. This paper shows that this is not the case. We establish the
capacity region for a class of 3-receiver broadcast channels with 2 degraded message 
sets and show that it can be strictly larger than the straightforward extension of 
the K\"{o}rner-Marton region. The key new idea is indirect decoding, whereby a
receiver who cannot directly decode a cloud center, finds it indirectly by decoding 
satellite codewords.  This idea is then used to establish new inner and outer bounds 
on the capacity region of the general 3-receiver broadcast channel with  2 and 3 
degraded message sets. We show that these bounds are tight for some nontrivial 
cases. The results suggest that the capacity of the 3-receiver broadcast channel 
with degraded message sets is as at least as hard to find as the capacity of the 
general 2-receiver broadcast channel with common and private message.
\end{abstract}


\section{Introduction}
\label{se:intro}
A broadcast channel with degraded message sets represents a scenario where a
sender wishes to communicate a common message to {\em all} receivers, a first
private message to a first subset of the receivers, a second private message to
a second subset of the first subset and so on. Such scenario can arise, for
example, in video or music broadcasting over a wireless network to nested
subsets of
receivers at varying levels of quality. The common message represents the lowest
quality version to be sent to all receivers, the first private message
represents the additional information needed for the first subset of 
receivers to decode the second lowest quality version, and so on. What is the
set of simultaneously achievable rates for communicating such degraded message
sets over the network?

This question was first studied by K\"{o}rner and Marton in
1977~\cite{kom77}. They considered a general 2-receiver discrete-memoryless
broadcast channel with sender $X$ and receivers $Y_1$ and $Y_2$. A common
message $M_0 \in [1,2^{nR_0}]$ is to be sent to both receivers and a
private message $M_1 \in [1,2^{nR_1}]$ is to be sent only to receiver
$Y_1$. They showed that the capacity region is given by the set of all
rate pairs $(R_0,R_1)$ such that~\footnote{The K\"{o}rner-Marton
characterization does not include the second term inside the min in the
first inequality, $I(U;Y_1)$. Instead it includes the bound $R_1+R_2 \le
I(X;Y_1)$. It can be easily shown that the two characterizations are
equivalent.}
\begin{align}
\label{eq:km}
R_0 &\le \min\{I(U;Y_1),I(U;Y_2)\},\\ \nonumber
R_1 &\le I(X;Y_1|U),
\end{align}
for some $p(u,x)$.  These rates are achieved using superposition
coding~\cite{cov72}. The common message is represented by the auxiliary
random variable $U$ and the private message is superimposed to
generate $X$. The main contribution of~\cite{kom77} is proving a
strong converse using the technique of images-of-a-set~\cite{kom77a}. 

Extending the K\"{o}rner-Marton result to more than 2 receivers has remained
open even for the simple case of 3 receivers $Y_1,Y_2,Y_3$ with 2 degraded
message sets, where a common
message $M_0$ is to be sent to all receivers and a private message
$M_1$ is to be sent only to receiver $Y_1$. The straightforward extension of the
K\"{o}rner-Marton region to this case yields the achievable rate region
consisting of the set of all rate pairs $(R_0,R_1)$ such that
\begin{align}
\label{eq:3-1KM}
R_0 &\le \min\{I(U;Y_1),I(U;Y_2),I(U;Y_3)\},\\ \nonumber
R_1 &\le I(X;Y_1|U),
\end{align}
for some $p(u,x)$. Is this region optimal?

\begin{figure}[ht]
\begin{center}
\begin{psfrags}
\psfrag{a}{$X$}
\psfrag{b}[c]{$p(y_1,y_2|x)$}
\psfrag{c}{$Y_1$}
\psfrag{d}[c]{$p(y_3|y_1)$}
\psfrag{e}{$Y_3$}
\psfrag{f}{$Y_2$}
\epsfig{figure=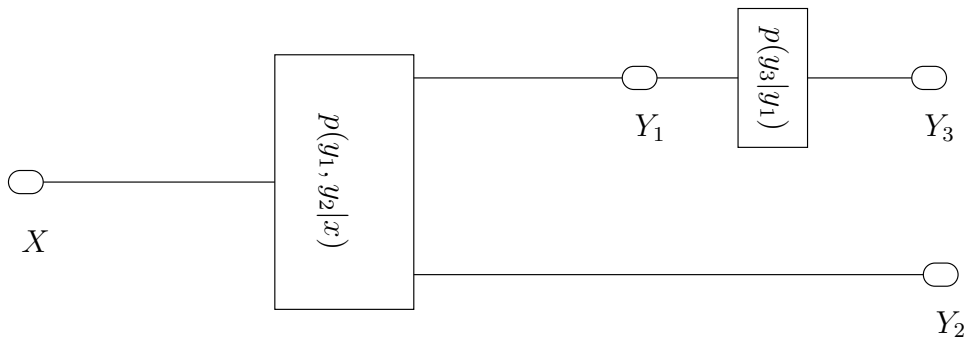,width=0.7\linewidth} \caption{Multilevel 3-receiver
broadcast channels. Message $M_0$ is to be sent to all receivers and message
$M_1$ is to be sent only to $Y_1$.} \label{fig:mult}
\end{psfrags}
\end{center}
\end{figure}

In~\cite{dit06}, it was shown that the above region (and its natural extension
to $k>3$
receivers) is optimal for a class of product discrete-memoryless and
Gaussian broadcast channels, where each of the receivers who decode only the
common 
message is a degraded version of the unique receiver that also decodes the
private message.
In~\cite{pdt07}, it was shown that a straightforward extension of
K\"{o}rner-Marton region 
is optimal for the class of linear deterministic broadcast channels, where the
operations are
performed in a finite field.  In addition to establishing the degraded message
set capacity for this class the authors gave an explicit characterization of the
optimal auxiliary random variables.
In a recent paper Borade et al.~\cite{bzt07} introduced {\em multilevel}
broadcast channels, which combine aspects of degraded broadcast channels and
broadcast channels with degraded message sets. They established an achievable
rate region as well as a ``mirror-image" outer bound for these channels. Their
achievable rate region is again a straightforward extension of the
K\"{o}rner-Marton region to
$k$-receiver multilevel broadcast channels. In particular,
Conjecture 5 of \cite{bzt07} states that the capacity region for the 3-receiver
multilevel broadcast
channels depicted in Figure~\ref{fig:mult} is  the set of all rate pairs
$(R_0,R_1)$ such that
\begin{align}
\label{eq:3-BZT}
R_0 &\le \min\{I(U;Y_2),I(U;Y_3)\},\\ \nonumber
R_1 &\le I(X;Y_1|U),
\end{align}
for some $p(u,x)$. Note that this region, henceforth referred to as {\em the BZT
region},  is the same as (\ref{eq:3-1KM}) because in the multilevel broadcast
channel $Y_3$ is a degraded version of $Y_1$ and therefore $I(U;Y_3) \le
I(U;Y_1)$. 

In this paper we show that the straightforward  extension of the
K\"{o}rner-Marton region
to more than 2 receivers is not in general optimal. We establish the capacity
region of the multilevel broadcast channels depicted in Figure~\ref{fig:mult}
as the set of rate pairs $(R_0,R_1)$ such that
\begin{align*}
R_0 &\le \min\{I(U_2;Y_2),I(U_1;Y_3) \},\\ \nonumber
R_0 + R_1 & \leq  \min \{ I(U_1;Y_3)+ I(X;Y_1|U_1), I(U_2;Y_2)+ I(X;Y_1|U_2)\},
\end{align*}
for some $p(u_1)p(u_2|u_1)p(x|u_2)$, and show that it can be strictly larger
than the BZT region.  In our coding scheme, the common message
$M_0$ is represented by $U_1$ (the cloud centers), part of $M_1$ is superimposed
on $U_1$ to obtain $U_2$ (satellite codewords), and the rest of $M_1$ is
superimposed on $U_2$ to yield $X$. Receivers $Y_1$ and $Y_3$ find $M_0$ by
decoding $U_1$, whereas receiver $Y_2$ who may not be able to directly decode
$U_1$, finds $M_0$ {\em indirectly} by
decoding a list of satellite codewords.  After decoding $U_1$, receiver
$Y_1$ finds $M_1$ by proceeding to decode $U_2$ then $X$.

The rest of the paper is organized as follows. In Section~\ref{sec:prelim}, we
provide needed definitions. In Section~\ref{se:capproof}, we establish the
capacity region for the multilevel broadcast channel in Figure~\ref{fig:mult}
(Theorem~\ref{th:cap}). In Section~\ref{se:bec}, we show through an example that
the capacity region for the multilevel broadcast channel  can be strictly
larger than the BZT region. In Section~\ref{se:ibobgendeg}, we extend the
results on the multilevel broadcast channel to establish inner and outer
bounds on the capacity region of the general 3-receiver broadcast channel with 2
degraded message sets
(Propositions~\ref{prop:innerbound} and \ref{prop:outerbound}). We show that
these bounds are tight when $Y_1$ is {\em less noisy} than $Y_3$
(Proposition~\ref{prop:caplessnoisy}), which is a more relaxed condition than
the degradedness condition of the multilevel broadcast channel model. We then
extend the inner bound to 3 degraded message sets (Theorem~\ref{th:geninner}).
Although Proposition~\ref{prop:innerbound} is a special case of
Theorem~\ref{th:geninner}, it is presented earlier for clarity of exposition.
Finally, we show that the inner bound of Theorem~\ref{th:geninner} when
specialized to the case
of 2 degraded message sets, where $M_0$ is to be sent to all receivers and $M_1$
is
to be sent to $Y_1$ and $Y_2$, reduces to the straightforward extension of the
K\"{o}rner-Marton region (Corollary~\ref{co:innerde2me}). We show that this
inner bound is tight for deterministic   broadcast channels (Proposition
\ref{prop:deterministic}) and when $Y_1$ is less noisy than $Y_3$ and $Y_2$ is
less noisy than $Y_3$ (Proposition~\ref{prop:lessnoisy2}). Finally, we outline a
general approach to obtaining inner bounds on capacity for general $k$-receiver
broadcast channel scenarios that uses the new idea of indirect decoding.

\section{Definitions}
\label{sec:prelim}
Consider a discrete-memoryless 3-receiver broadcast channel consisting of  an
input alphabet $\mathcal{X}$, output alphabets $\mathcal{Y}_1$, $\mathcal{Y}_2$ 
and
$\mathcal{Y}_3$, and a probability transition function $p(y_1,y_2,y_3|x)$. 

A $(2^{nR_0},2^{nR_1},n)$ {\em 2-degraded message set} code for a 3-receiver
broadcast channel consists
of (i) a pair of messages $(M_0,M_1)$ uniformly distributed over
$[1,2^{nR_0}]\times [1, 2^{nR_1}]$, (ii) an
encoder that assigns a codeword $x^n(m_0,m_1)$, for each message
pair $(m_0,m_1) \in [1, 2^{nR_0}]\times
[1,2^{nR_1}]$, and (iii) three decoders, one that maps each received
$y_1^n$ sequence into an estimate $(\hat m_{01},\hat m_1)\in
[1,2^{nR_0}]\times [1, 2^{nR_1}]$, a second that maps each
received $y_2^n$ sequence into an estimate $\hat{m}_{02}\in  [1, 2^{nR_0}]$,
and a third that maps each
received $y_3^n$ sequence into an estimate $\hat{m}_{03}\in  [1, 2^{nR_0}]$.

The probability of error is defined as
\begin{equation*}
P_e^{(n)} =  \Prob\{ \hat{M}_1\neq M_1 \; \text{\rm or } \hat{M}_{0k}\neq M_{0}
\text{ for $k=1,2,$ or } 3\}.
\end{equation*}
A rate tuple $(R_0,R_1)$ is said to be achievable if there exists
a sequence of $(2^{nR_0}, 2^{nR_1},n)$ 2-degraded message set codes with
$P_e^{(n)} \to 0$. The capacity region of the broadcast
channel is the closure of the set of
achievable rates. 

A 3-receiver {\em multilevel} broadcast channel~\cite{bzt07} is a 3-receiver
broadcast
channel with 2 degraded message sets where $p(y_1,y_2,y_3|x)=
p(y_1,y_2|x)p(y_3|y_1)$ for every $(x,y_1,y_2,y_3) \in \mathcal{X} \times
\mathcal{Y}_1 \times \mathcal{Y}_2 \times \mathcal{Y}_3$ (see
Figure~\ref{fig:mult}).  

In addition to considering the multilevel 3-receiver broadcast channel and the
general 3-receiver broadcast channel with 2 degraded message sets, we shall also
consider the following two scenarios: 
\begin{enumerate}
\item 3-receiver broadcast channel with 3 message sets, where $M_0$ is to be
sent to all receivers, $M_1$ is to be sent to $Y_1$ and $Y_2$, and $M_2$ is to
be sent only to $Y_1$.
\item 3-receiver broadcast channel with 2 degraded message sets, where $M_0$ is
to be
sent all receivers and $M_1$ is to be sent to both $Y_1$ and $Y_2$. 
\end{enumerate}
Definitions
of codes, achievability and capacity regions for these cases are straightforward
extensions of the above definitions. Clearly, the 2 degraded message set
scenarios are special cases of the 3 degraded message set. Nevertheless, we
shall begin with the special class of multilevel broadcast channel because we
are able to establish its capacity region.

\section{Capacity of 3-Receiver Multilevel Broadcast Channel}
\label{se:capproof}

The key result of this paper is given in the following theorem.
\medskip

\begin{theorem}
\label{th:cap}
The capacity region of the 3-receiver multilevel broadcast channel in
Figure~\ref{fig:mult} is the set of rate pairs $(R_1,R_2)$ such that
\begin{align}
\label{eq:capacity}
R_0 & \leq  \min \{ I(U_1;Y_3), I(U_2;Y_2)\}, \\ \nonumber
R_0 + R_1 & \leq  \min \{I(U_1;Y_3) + I(X;Y_1|U_1), I(U_2;Y_2) + I(X;Y_1|U_2), 
\}.
\end{align}
for some $p(u_1)p(u_2|u_1)p(x|u_2)$, where the cardinalities of the auxiliary
random variables satisfy $\|\mathcal{U}_1\| \leq \|\mathcal{X}\|+4$ and $\|U_2\|
\leq \|
\mathcal{X} \|^2 + 5\|\mathcal{X}\|+4$.
\end{theorem}
\medskip

\noindent{\em Remarks:}
\begin{enumerate}
\item It is easy to show by setting $U_1 = U_2 = U$ in the above theorem that
the BZT region~(\ref{eq:3-BZT}) is contained in the capacity
region~\eqref{eq:capacity}. We show in the
next section that the capacity region
(\ref{eq:capacity}) can be strictly larger the BZT region.
\item It is straightforward to show that the  above region is convex and
therefore there is no need to use a time-sharing auxiliary random variable.
\end{enumerate}
The proof of Theorem~\ref{th:cap} is given in the following  subsections. 
\subsection{Converse}
\label{sse:outerbound}
We show that the region in Theorem
\ref{th:cap} forms an outer bound to the capacity region. 
The proof is quite similar to previous weak converse and outer bound proofs for
2-receiver
broadcast channels (e.g., see \cite{gal74,elg79,nae07}). Suppose we are given a
sequence of codes for the multilevel broadcast channel with $P_e^{(n)} \to 0$.
For each code, we form the empirical distribution for $M_0,M_1,X^n$. 

Since $X \to Y_1 \to Y_3$ forms a {\em physically degraded} broadcast
channel, it follows that the code rate pair $(R_0,R_1)$ must satisfy the
inequalities
\begin{align*}
R_0 & \leq  I(U_1;Y_3),\\
R_1 & \leq  I(X;Y_1|U_1), 
\end{align*}
for some $p(u_1,x)$~\cite{gal74}, where $U_1,X,Y_1,Y_3$ are defined as
follows.  
Let $U_{1i} =(M_0,Y_1^{i-1}),\ i = 1,\ldots,n$, and  let
$Q$ be a time-sharing random variable uniformly distributed over the set
$\{1,2,...,n\}$ and  independent of $X^n,Y_1^n,Y_2^n,Y_3^n$. We then set $U_1 =
(Q,U_{1Q})$ and $X = X_Q$, $Y_1 = Y_{1Q}$, and $Y_3 = Y_{3Q}$. Thus, we have
shown that
\begin{align*}
R_0 & \leq  I(U_1;Y_3),\\
R_0+R_1 & \leq  I(U_1;Y_3)+I(X;Y_1|U_1). 
\end{align*}

Next, since the decoding requirements of the broadcast channel $X
\to (Y_1,Y_2)$ makes it a broadcast channel with degraded message sets, the code
rate pair must satisfy the  inequalities
\begin{align*}
R_0 & \leq  \min \{I(U_2;Y_2), I(U_2,Y_1)\},\\
R_0 + R_1 & \leq   I(U_2;Y_2)+ I(X;Y_1|U_2),
\end{align*}
for some $p(u_2,x)$~\cite{elg79}, where $U_2$ is identified as follows. Let
$U_{2i} =(M_0,Y_1^{i-1},Y_{2~i+1}^{~~n}),\ i = 1,\ldots,n$, then we set
$U_2=(Q,U_{2Q})$. 

Combining the above two outer bounds, we see that $U_1 \to U_2 \to X$ forms a
Markov chain. Observe that this Markov nature of the auxiliary random variables
along with the degraded nature of $X \to Y_1 \to Y_3$  implies that $I(U_2;Y_1)
\geq I(U_2;Y_3) \geq I(U_1;Y_3)$. 

Thus we have shown that the code rate pair $(R_0,R_1)$ must satisfy the
inequalities
\begin{align*}
R_0 & \leq  \min \{ I(U_1;Y_3), I(U_2;Y_2) \}, \\
R_0 + R_1 & \leq \min \{ I(U_1;Y_3)+ I(X;Y_1|U_1), I(U_2;Y_2) + I(X;Y_1|U_2)
\},
\end{align*}
for some  $p(u_1)p(u_2|u_1)p(x|u_2)$. This establishes the converse to Theorem
~\ref{th:cap}.

\subsection{Achievability}
\label{sse:ib}
The interesting part of the proof of Theorem ~\ref{th:cap} is
achievability. Specifically, step 3  of the decoding procedure for Case 2 below
describes a key contribution of this paper. We show how $Y_2$ can find $M_0$
without directly decoding $U_1^n$ or uniquely decoding $U_2^n$. 

To show achievability of any rate pair $(R_0,R_1)$ in region
(\ref{eq:capacity}),
because of its convexity, it suffices to show the achievability of any rate
pair $(R_0,R_1)$ such that
\begin{align*}
R_0 &= \min \{ I(U_1;Y_3), I(U_2;Y_2) \} -\delta \\
R_0+R_1 &= \min \{ I(U_1;Y_3)+ I(X;Y_1|U_1), I(U_2,Y_2)+ I(X;Y_1|U_2) \}
-3\delta,
\end{align*}
for some $U_1 \to U_2 \to X$ and any $\delta >0$.

Rewriting the second inequality we obtain
\begin{align*}
R_0 + R_1 &= I(U_1;Y_3) + \min\{I(U_2;Y_1|U_1),I(U_2;Y_3)-I(U_1;Y_3)\} +
I(X;Y_1|U_2) - 3\delta.
\end{align*}

Now consider the following two cases:
\medskip

\noindent{\em Case 1:} $I(U_1;Y_3) > I(U_2;Y_2)$: The rates reduce to
\begin{align*}
R_0&= I(U_2;Y_2) -\delta\\
R_1&= I(X;Y_1|U_2) -2\delta.
\end{align*}
This pair can be achieved via a simple superposition coding scheme~\cite{cov72}.
\medskip

\noindent{\em Case 2:} $I(U_1;Y_3) \le I(U_2;Y_2)$: The rates reduce to
\begin{align*}
R_0&= I(U_1;Y_3) -\delta\\
R_1&= I(X;Y_1|U_2) + \min \{ I(U_2;Y_1|U_1),  I(U_2,Y_2)-I(U_1;Y_3)\}
-2\delta.
\end{align*}
Let $S_{1} = \min \{ I(U_2;Y_1|U_1),  I(U_2,Y_2)-I(U_1;Y_3)\} -\delta$ and
$S_{2}=  I(X;Y_1|U_2) - \delta$, then $R_1 = S_{1} +S_{2}$.
\medskip

\noindent{\em Code Generation:}  
\smallskip

Fix $p(u_1)p(u_2|u_1)p(x|u_2)$ that satisfies the condition of Case 2. Generate
$2^{nR_0}=2^{n(I(U_1;Y_3) - \delta)}$  sequences $U_1^n(1),\ldots,
U_1^n(2^{nR_0})$ distributed uniformly at random over the set of
$\epsilon$-typical\symbolfootnote[2]{We assume strong typicality throughout this
paper~\cite{cot91}.} $U_1^n$
sequences, where $\delta \to 0$ as $\epsilon \to 0$. For each $U_1^n(m_0)$,
generate $2^{nS_1} = 2^{n(\min
\{I(U_2;Y_1|U_1), I(U_2,Y_2)-I(U_1;Y_3)\} -\delta)}$  sequences 
$U_2^n(m_0,1),$ $U_2^n(m_0,2)$, $ \ldots$, $U_2^n(m_0,2^{nS_1})$ distributed
uniformly at random over the set of conditionally $\epsilon$-typical $U_2^n$
sequences. For
each $U_2^n(m_0,s_{1})$ generate $2^{nS_2} =2^{n(I(X;Y_1|U_2) - \delta)}$
sequences
$X^n(m_0,s_{1},1)$, $X^n(m_0,s_{1},2)$, $\ldots$, $X^n(m_0,s_{1},2^{nS_2})$
distributed
uniformly at random over the set of conditionally $\epsilon$-typical $X^n$
sequences. 

\medskip
\noindent{\em Encoding:}

To send the message pair $(m_0,m_1) \in [1,2^{nR_0}]\times  [1,2^{nR_1}]$, the
sender expresses $m_1$ by the pair $(s_1,s_2) \in  [1,2^{nS_1}]\times 
[1,2^{nS_2}]$ and sends $X^n(m_0,s_1,s_2)$.
\medskip

\noindent{\em Decoding and Analysis of Error Probability:} 
\begin{enumerate}
\item Receiver $Y_3$ declares that $m_0$ is sent if it is the unique message
such that $U_1^n(m_0)$ and $Y_3^n$ are jointly $\epsilon$-typical.
It is easy to see that this can be achieved with arbitrarily small probability
of error because $R_0 = I(U_1;Y_3) - \delta$. 

\item Receiver $Y_1$ first declares that $m_0$ is sent if it is the unique
message such that $U_1^n(m_0)$ and $Y_1^n$ are jointly $\epsilon$-typical. This
decoding
step succeeds with arbitrarily high probability because $R_0 = I(U_1;Y_3)
-\delta
\leq I(U_1;Y_1) - \delta$. It then declares 
that $s_1$ is sent if it is the unique index such
that $U_2^n(m_0,s_1)$ and $Y_1^n$ are jointly $\epsilon$-typical. This decoding
step
succeeds with arbitrarily high probability because $S_1 \leq I(U_2;Y_1|U_1)-
\delta$. Finally it declares that $s_2$  is sent if it is the unique index
such that $X^n(m_0,s_1,s_2)$ and $Y_1^n$ are jointly $\epsilon$-typical.  This
decoding
step succeeds with high probability because $S_2 = I(X;Y_1|U_2) - \delta$.

\item Receiver $Y_2$ finds $M_0$ as follows. It declares that $m_0\in
[1,2^{nR_0}]$ is sent if it is the unique index such that $U_2^n(m_0,s_1)$ and
$Y_2^n$ are jointly $\epsilon$-typical for some $s_1 \in [1,2^{nS_1}]$. Suppose
$(1,1)\in
[1,2^{nR_0}]\times [1,2^{nS_1}]$ is the message pair sent, then the probability
of error averaged over the choice of codebooks can be upper bounded as follows
\begin{align*}
P_e^{(n)} &\le \Prob\{ (U_2^n(1,1),Y_2^n)\text{ are not jointly
$\epsilon$-typical}\} \\
&\qquad +  \Prob\{ (U_2^n(m_0,s_1),Y_2^n)\text{ are jointly $\epsilon$-typical
for some } m_0\neq 1\}\\
&\stackrel{(a)}{<} \delta' + 2^{n(R_0+S_1)} \sum_{m_0\neq
1}\sum_{s_1}\Prob\{(U_2^n(m_0,s_1),Y_2^n) \text{ jointly $\epsilon$-typical}\}\\
&\stackrel{(b)}{\le} \delta' + 2^{n(R_0+S_1)} 2^{-n(I(U_2;Y_2)-\delta)}\\
&\stackrel{(c)}{\le} \delta' + 2^{-n\delta},
\end{align*}
where $(a)$ follows by the union of events bound, $(b)$ follows by the fact that
for $m_0\neq 1$, $U_2^n(m_0,s_1)$ and $Y_2^n$ are  generated completely
independently and thus each probability term under the sum is upper bounded by
$2^{-n(I(U_2;Y_2)-\delta)}$~\cite{cot91}, $(c)$ follows because by construction 
$R_0 + S_1 \leq I(U_2;Y_2)-2\delta$, $\delta'\to 0$ as $\epsilon \to 0$. Thus
with arbitrarily high probability, any jointly $\epsilon$-typical
$U_2^n(m_0,s_1)$ with the received $Y_2^n$ sequence must be of the form
$U_2^n(1,s_1)$, and receiver $Y_2$ can
correctly decodes $M_0$ with arbitrarily small probability of error. Note that
$Y_2$ may or may not be able to uniquely decode $U^n_2(1,1)$. However, it finds
the correct common message with arbitrarily small probability of error even if
its rate $R_0 > I(U_1;Y_2)$! 

\medskip

\end{enumerate}

Thus all receivers can decode their intended messages with arbitrarily small
probability of error and hence the rate pair $R_0 = I(U_1;Y_3) - \delta, R_1 =
I(X;Y_1|U_2) + \min \{ I(U_2;Y_1|U_1),  I(U_2,Y_2)-I(U_1;Y_3)\} -2\delta$ is
achievable. This completes the proof of achievability of Theorem \ref{th:cap}.
\medskip

\noindent{Remarks:} 
\begin{enumerate}
\item We denote the decoding technique used in step 3 as {\em indirect
decoding}, since $Y_2$ decodes the cloud center $U_1$ indirectly by decoding
satellite codewords.
\item There is no need to break up the coding scheme into two cases. The coding
scheme for Case 2 suffices. This will become clear when we prove the achievable
region for the general case of 3-receivers with 2 degraded message sets in
Proposition~\ref{prop:innerbound}.
\end{enumerate}
\medskip

\subsection{Bounds on Cardinality}
\label{sse:card}
Using the strengthened Carath\'{e}odory theorem by Fenchel and Eggleston
\cite{czk78} it can be readily shown that for any choice of the auxiliary random
variable $U_1$, there exists  a random variable $V_1$ with cardinality bounded
by $\| \mathcal{X} \| + 1$  such that $I(U_1;Y_3) = I(V_1;Y_3)$ and
$I(X;Y_1|U_1) = I(X;Y_1|V_1)$.  Similarly for any choice of $U_2$, one can
obtain a random variable $V_2$ with cardinality bounded by $\| \mathcal{X} \| +
1$ such that $I(U_2;Y_2) = I(V_2;Y_2)$ and $I(X;Y_1|U_2) = I(X;Y_1|V_2)$. While
this preserves the region, it is not clear that the new random variables $V_1,
V_2$ would preserve the Markov condition $V_1 \to V_2 \to X$. To circumvent this
problem we adapt arguments from \cite{czk78} to establish the cardinality bounds
stated in Theorem \ref{th:cap}. For completeness, we provide an outline of the
argument. 

Given $U_1 \to U_2 \to X \to (Y_1,Y_2, Y_3)$, we need to show the existence of
random
variables $V_1, V_2$ such that the following conditions hold: $V_1 \to V_2 \to
X$ forms a
Markov chain, $I(V_1;Y_3)=I(U_1;Y_3)$, $I(V_2;Y_2)=I(U_2;Y_2)$,
$I(X;Y_1|V_1)=I(X;Y_1|V_1)$, and $I(X;Y_1|V_2)=I(X;Y_1|U_2)$. Further, 
the cardinalities of the new random variables must satisfy $\|\mathcal{V}_1\|
\leq \| \mathcal{X} \| + 4$, $\|\mathcal{V}_2\| \leq \|\mathcal{X}\|^2 + 5
\|\mathcal{X}\| + 4$. 

This argument is proved in two steps. In the first step a random variable $V_1$
and  transition probabilities $p(u_2|v_1)$ are constructed such that the
following are held constant: $p(x)$, the marginal probability of $X$ (and hence
$Y_1, Y_2, Y_3$), $H(Y_1|U_1)$, $H(Y_2|U_1)$, $H(Y_3|U_1)$, $H(Y_2|U_2,U_1)$,
and $H(Y_1|U_2,U_1)$.
Using standard arguments \cite{ahk75,czk78}, there exists a random variable
$V_1$ and transition probabilities $p(u_2|v_1)$, with cardinality of $V_1$
bounded by $\| \mathcal{X} \| + 4$,
such that the above equalities are achieved. In particular the marginals of
$X,Y_1,Y_2,Y_3$ are held constant. However the distribution of $U_2$ is not
necessarily held constant and hence we shall denote the resulting random
variable as $U_2'$. 

We thus have random variables $V_1 \to U_2' \to X$ such that 
\begin{align}
I(V_1;Y_3) &= I(U_1;Y_3), \nonumber \\
I(V_1;Y_2) &= I(U_1;Y_2), \nonumber \\
I(X;Y_1|V_1) &= I(X;Y_1|U_1), \label{eq:firststage} \\
I(U_2';Y_1|V_1) &= I(U_2; Y_1|U_1).\nonumber
\end{align}

In the second step, for each $V_1=v_1$ a new random variable $V_2(v_1)$ is found
such that the following are held constant: $p(x|v_1)$, the marginal distribution
of $X$ conditioned on $V_1 = v_1$, $H(Y_1|U_2',V_1=v_1)$, and
$H(Y_2|U_2',V_1=v_1)$. Again  standard arguments imply that there exists a
random variable $V_2(v_1)$ and transition probabilities $p(x|v_2(v_1))$, with
cardinality of $V_1$ bounded by $\| \mathcal{X} \| + 1$, such that the above
equalities are achieved. This in particular implies that
\begin{align} 
I(V_2(V_1);Y_2|V_1) &= I(U_2'; Y_2|V_1) = I(U_2;Y_2|U_1), \label{eq:secondstage}
\\
I(V_2(V_1);Y_1|V_1) &= I(U_2';Y_1|V_1) = I(U_2; Y_1|U_1).\nonumber 
\end{align}

Now, set $V_2 = (V_1,V_2(V_1))$ and observe the following as a consequence of
Equations \eqref{eq:firststage} and \eqref{eq:secondstage}.
\begin{align*}
I(V_2;Y_2) &= I(V_1;Y_2) + I(V_2(V_1);Y_2|V_1) = I(U_1;Y_2) + I(U_2;Y_2|U_1) =
I(U_2;Y_2), \\
I(X;Y_1|V_2) &= I(X;Y_1|V_1) - I(V_2(V_1);Y_1|V_1) = I(X;Y_1|U_1) -
I(U_2;Y_1|U_1) = I(X:Y_1|U_2).
\end{align*}
We thus have the required random variables $V_1,V_2$ satisfying the cardinality
bounds $\| \mathcal{X} \|+4 $ and $(\| \mathcal{X} \|+4)(\| \mathcal{X} \|+1)$,
respectively as desired.

\section{Multilevel Product Broadcast Channel} 
\label{se:bec}
In this section we show that the BZT region can be strictly smaller than the
capacity region in Theorem~\ref{th:cap}. 

Consider the product of 3-receiver broadcast channels given by the Markov
relationships
\begin{align}
\label{eq:channel}
X_1 & \to Y_{21} \to Y_{11} \to Y_{31}, \nonumber \\
X_2 & \to Y_{12} \to Y_{32}.
\end{align}
In Appendix~\ref{se:append-bzt} we derive the following simplified
characterizations for the
capacity and the BZT regions. 
\medskip

\begin{proposition}
\label{prop:char-bzt}
The BZT region for the above product channel reduces to the set of
rate pairs $(R_0,R_1)$ such that
\begin{subequations}
\label{eq:PBZT}
\begin{align}
R_0 &\le I(V_1; Y_{31}) + I(V_2; Y_{32}), \label{eq:bzt-term1}\\ 
R_0 &\le I(V_1; Y_{21}), \label{eq:bzt-term2}\\ 
R_1 &\le I(X_1;Y_{11}|V_1) + I(X_2; Y_{12} |V_2), \label{eq:bzt-term3}
\end{align}
\end{subequations}
for some $p(v_1)p(v_2) p(x_1|v_1) p(x_2|v_2)$.
\end{proposition}
\medskip

\begin{proposition}
\label{prop:char-cap}
The capacity region for the product channel reduces
to the set of rate
pairs $(R_0,R_1)$ such that
\begin{subequations}
\label{eq:pcapacity}
\begin{align}
R_0 &\le I(V_{11}; Y_{31}) + I(V_{12}; Y_{32}), \label{eq:cap-term1}\\ 
R_0 &\le I(V_{21}; Y_{21}), \label{eq:cap-term2}\\ 
R_0+R_1 &\le I(V_{11}; Y_{31}) + I(V_{12}; Y_{32}) + I(X_1;Y_{11}|V_{11}) +
I(X_2;Y_{12} |V_{12}),\label{eq:cap-term3} \\ 
R_0+R_1 & \le I(V_{21}; Y_{21}) + I(X_1;Y_{11}|V_{21}) + I(X_2; Y_{12} |V_{12}),
\label{eq:cap-term4}
\end{align}
\end{subequations}
for some $p(v_{11})p(v_{21}|v_{11}) p(x_1|v_{21}) p(v_{12}) p(x_2|v_{12})$.
\end{proposition}
\medskip

Now we compare these two regions via the following example.

\medskip

\subsection*{Example:}
\label{sse:example}

\begin{figure}[ht]
\small
\begin{center}
\begin{psfrags}
\psfrag{a}{$X_1$}
\psfrag{b}{$Y_{21}$}
\psfrag{c}{$Y_{11}$}
\psfrag{d}{$Y_{31}$}
\psfrag{e}{$X_2$}
\psfrag{f}{$Y_{12}$}
\psfrag{g}{$Y_{32}$}
\psfrag{x}[r]{$1/2$}
\psfrag{y}[rb]{$1/2$}
\psfrag{u}[r]{$2/3$}
\psfrag{v}[rb]{$2/3$}
\psfrag{s}[r]{$1/2$}
\psfrag{t}[rb]{$1/2$}
\psfrag{p}[b]{$1/2$}
\psfrag{q}[t]{$1/2$}
\psfrag{l}[b]{$1/3$}
\psfrag{m}[t]{$1/3$}
\psfrag{n}[b]{$1/2$}
\psfrag{o}[t]{$1/2$}
\psfrag{h}[r]{$0$}
\psfrag{i}[r]{$1$}
\psfrag{j}[r]{$0$}
\psfrag{k}[r]{$1$}
\psfrag{A}[l]{$0$}
\psfrag{B}[l]{$E$}
\psfrag{C}[l]{$1$}
\psfrag{D}[l]{$0$}
\psfrag{E}[l]{$E$}
\psfrag{F}[l]{$1$}
\epsfig{figure=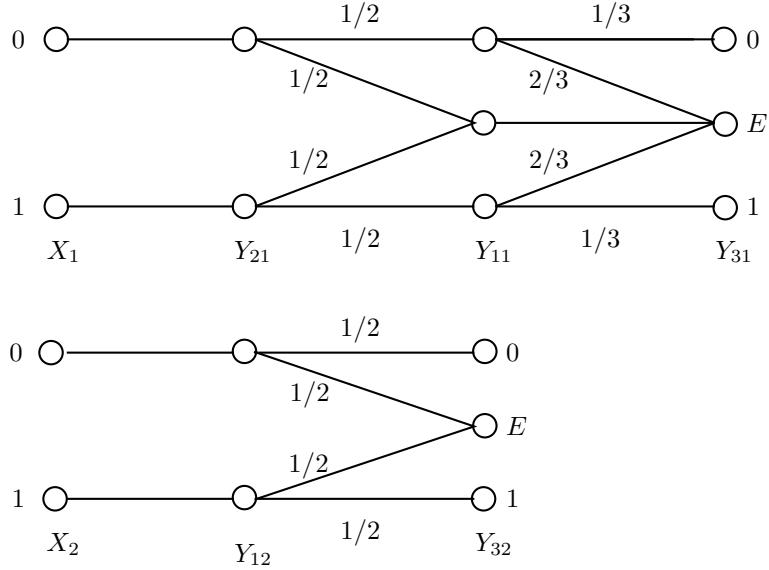,width=0.55\linewidth} \caption{Product multilevel
broadcast channel example.} \label{fig:bec}
\end{psfrags}
\end{center}
\end{figure}

Consider the multilevel product broadcast channel example in Figure
\ref{fig:bec}, where: $\mathcal{X}_1=\mathcal{X}_2=\mathcal{Y}_{12}=
\mathcal{Y}_{21}= \{0,1\}$, and $
\mathcal{Y}_{11}=\mathcal{Y}_{31}=\mathcal{Y}_{32} = \{0,E,1\}$, $Y_{21} = X_1$,
$Y_{12} = X_2$, 
the channels $Y_{21} \to Y_{11}$ and $Y_{12} \to Y_{32}$ are binary erasure
channels (BEC)
with erasure probability $\frac 12$, and  the channel $Y_{11} \to Y_{31}$ is
given by the transition probabilities: $\Prob\{Y_{31}=E|Y_{11} = E\} = 1,\
\Prob\{Y_{31}=E|Y_{11} =
0\} = \Prob\{Y_{31}=E|Y_{11} = 1\} =  2/3,\ \Prob(Y_{31}=0|Y_{11} =
0\} = \Prob\{Y_{31}=1|Y_{11} = 1\} =  1/3$. Therefore, the channel $X_1 \to
Y_{31}$ is effectively a BEC with erasure probability $5/6$.

The BZT region can be simplified to the following.
\medskip

\begin{proposition}
\label{prop:eval-bzt}
The BZT region for the above example reduces to the set of rate pairs $(R_0,
R_1)$
satisfying
\begin{align}
R_0 & \leq \min \left\{ \frac p6 + \frac q2, p\right\} , \nonumber \\
\label{eq:peval-bzt}
R_1 & \leq \frac{1-p}{2} + 1-q.
\end{align}
for some $0 \leq p,q \leq 1$.
\end{proposition}

The proof of this proposition is given in Appendix \ref{se:append-bzt}.
It is quite straightforward to see that $(R_0, R_1) = (\frac 12, \frac{5}{12})$
lies on the boundary of this region.

\medskip

The capacity region can be simplified to the following

\medskip
\begin{proposition}
\label{prop:eval-cap}
The capacity region for the channel in Figure \ref{fig:bec} reduces to set of
rate pairs $(R_0, R_1)$ satisfying
\begin{align}
R_0 & \leq \min \left\{ \frac r6 + \frac s2,  t\right\}, \nonumber \\
\label{eq:peval-cap}
R_0 + R_1 & \leq \min \left\{\frac r6 + \frac s2 + \frac{1-r}{2} + 1-s, t +
\frac{1-t}{2} + 1-s\right\},
\end{align}
for some $0 \leq r \leq t \leq 1, 0 \leq s \leq 1$.
\end{proposition}
\medskip

The proof of this proposition is also given in Appendix \ref{se:append-bzt}.
Note that substituting $r=t$ yields the BZT region.
By setting $r=0, s=1, t=1$ it is easy to see that $(R_0, R_1) = (1/2, 1/2)$ lies
on the boundary of the capacity region. On the other hand, for
$R_0=1/2$, the maximum achievable $R_1$ in the BZT region
is $5/12$. Thus the capacity region is strictly larger than the BZT
region.

\begin{figure}[htpb]
\begin{center}
\begin{psfrags}
\psfrag{R1}[b]{$R_1$}
\psfrag{R0}[t]{$R_0$}
\epsfig{figure=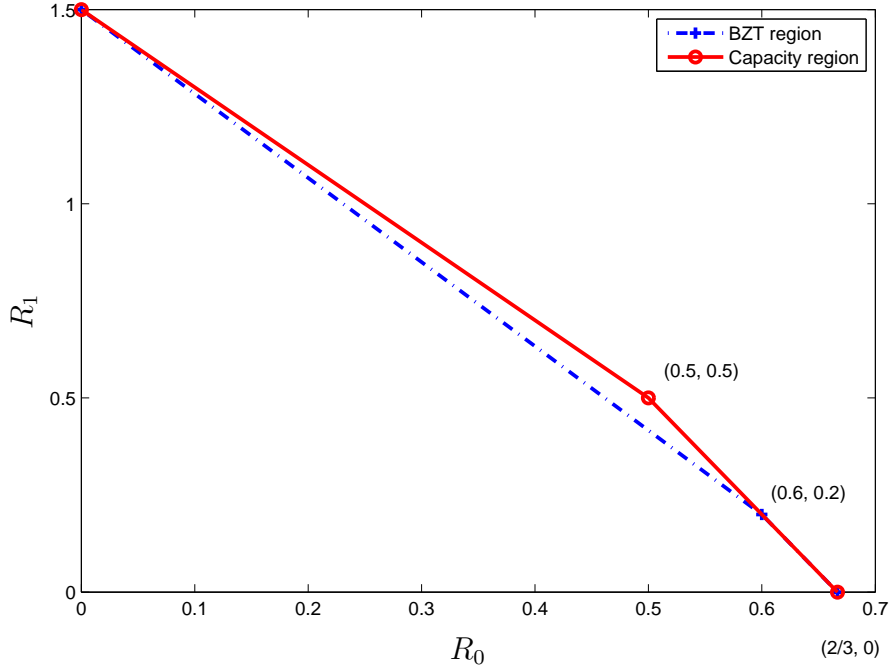,width=0.75\linewidth} 
\caption{The BZT and the capacity regions for the channel in
Figure~\ref{fig:bec}.} 
\label{fig:bztcap}
\end{psfrags}
\end{center}
\end{figure}

Figure \ref{fig:bztcap} plots the BZT region and the capacity region for the
example
channel. Both regions are specified by two line segments. The boundary of the
BZT
regions consists of the line segments: $(0,3/2)$ to
$(0.6,0.2)$ and $(0.6,0.2)$ to $(2/3,0)$. The capacity
region on the other hand is formed by the pair of line segments: $(0,3/2)$ to
$(1/2,1/2)$ and $(1/2,1/2)$ to $(2/3, 0)$. Note
that the boundaries of the two regions coincide on the line segment joining
$(0.6,0.2)$ to $(2/3,0)$.


\medskip

\noindent{\em Remarks:}
\begin{enumerate}
\item  Consider a 3-receiver Gaussian product multilevel broadcast channel,
where
\begin{align*}
Y_{21} &= X_1 + Z_1,\; 
Y_{11} = Y_{21} + Z_2,\; 
Y_{31} = Y_{11} + Z_3,\\
Y_{22} &= X_2 + Z_4,\;
Y_{32} = Y_{22} + Z_5.
\end{align*}
The noise power for $Z_i$ is $N_i$ for $i=1,2,\ldots,5$. We assume a total
average power constraint $P$ on $X=(X_1,X_2)$. 

Using Gaussian signalling it can be easily shown that the BZT region is the set
of all $(R_0,R_1)$ such that
\begin{align}
R_0 &\le {\cal C}\left( \frac{\alpha P_1}{ \bar \alpha P_1 + N_1 +N_2
+N_3}\right) + {\cal C}\left( \frac{\beta (P-P_1)}{ \bar \beta (P-P_1) + N_4
+N_5}\right), \\ \nonumber
R_0 &\le {\cal C}\left( \frac{\alpha P_1}{ \bar \alpha P_1 + N_1}\right), \\
\nonumber
R_1 &\le {\cal C}\left( \frac{ \bar \alpha P_1}{ N_1 +N_2}\right) + {\cal
C}\left( \frac{ \bar \beta (P-P_1)}{ N_4}\right),
\end{align}
for some $0 \le P_1 \le P$, $0 \le \alpha, \beta \le 1$.

Now if we use Gaussian signaling to evaluate region \eqref{eq:pcapacity},  one
obtains the achievable rate region consisting of the set of all $(R_0,R_1)$
such that
\begin{align}
R_0 &\le {\cal C}\left( \frac{a_1 P_1}{ \bar a_1 P_1 + N_1 +N_2 +N_3}\right) +
{\cal C}\left( \frac{a_2 (P-P_1)}{ \bar a_2 (P-P_1) + N_4
+N_5}\right),\nonumber \\ \nonumber
R_0 &\le {\cal C}\left( \frac{(a_1+b_1) P_1}{(1-a_1-b_1) P_1 + N_1}\right), \\
\nonumber
R_0+ R_1 &\le {\cal C}\left( \frac{\bar a_1 P_1}{ N_1 +N_2}\right) + {\cal
C}\left( \frac{ \bar a_2 (P-P_1)}{ N_4}\right) + {\cal C}\left( \frac{a_1 P_1}{
\bar a_1 P_1 + N_1 +N_2 +N_3}\right)\\ \label{eq:gaucap}
& \quad +
{\cal C}\left( \frac{a_2 (P-P_1)}{ \bar a_2 (P-P_1) + N_4 +N_5}\right), \\
\nonumber
R_0+ R_1 &\le {\cal C}\left( \frac{((1-a_1-b_1) P_1}{ N_1 +N_2}\right) + {\cal
C}\left( \frac{ (1-a_2-b_2) (P-P_1)}{ N_4}\right)\\ \nonumber
& \quad +  {\cal C}\left( \frac{(a_1+b_1) P_1}{(1-a_1-b_1) P_1 + N_1}\right), 
\end{align}
for some $0 \le P_1 \le P$, $0 \le a_1, a_2, b_1, b_2, a_1+b_1, a_2+b_2 \le 1$.

Now consider the above regions with the parameters values: $P=1, N_1=0.4,
N_2=N_3=0.1, N_4=0.5, N_5=0.1$. Fixing  $R_1=0.5 \log (0.49/0.3)$, we can show
that
the maximum achievable $R_0$ in the Gaussian BZT region is $0.5 \log
(2.0566...)$. On the other hand, using the parameter values $b_1=0.05,
1-a_1 = 0.1725$, $ 1-a_2 = 0.5079$, and $P_1 = 0.5680$ for the region given
by~\eqref{eq:gaucap},   the pair $(0.5 \log(2.0603), 0.5 \log(0.49/0.3))$ is in
the exterior of the region. Thus restricted to Gaussian signalling the BZT
region~\eqref{eq:PBZT} 
is strictly contained in region~\eqref{eq:pcapacity}. However, we have not been
able to prove that
Gaussian signaling
is optimal for either the BZT region or the capacity region. 

\item The reader may ask why we did not consider the more general product
channel
\begin{align*}
X_1 & \to Y_{21} \to Y_{11} \to Y_{31}, \\
X_2 & \to Y_{12} \to Y_{32} \to Y_{22}.
\end{align*}

In fact we considered this more general class at first but were unable to
show that the capacity region conditions reduce to the separated form
\begin{align*}
R_0 &\le I(V_{11}; Y_{31}) + I(V_{12}; Y_{32}),\\ \nonumber
R_0 &\le I(V_{21}; Y_{21}) + I(V_{22}; Y_{22}),\\ \nonumber
R_0+R_1 &\le I(V_{11}; Y_{31}) + I(V_{12}; Y_{32}) + I(X_1;Y_{11}|V_{11}) +
I(X_2;
Y_{12} |V_{12}),\\ \nonumber
R_0+R_1 & \le I(V_{21}; Y_{21})  + I(V_{22}; Y_{22}) + I(X_1;Y_{11}|V_{21}) +
I(X_2; Y_{12} |V_{12}),
\end{align*}
for some $p(v_{11})p(v_{21}|v_{11}) p(x_1|v_{21}) p(v_{12})
p(v_{22}|v_{12})p(x_2|v_{22})$.
\end{enumerate}

\section{Bounds on Capacity of General 3-Receiver Broadcast Channel with
Degraded Message Sets}
\label{se:ibobgendeg}

In this section we extend the results in Section~\ref{se:capproof} to obtain
inner and outer bounds on the capacity region of general 3-receiver broadcast
channel with degraded message sets. We first consider the same 2 degraded
message set
scenario as in Section~\ref{se:capproof} but without the condition that $X \to
Y_1 \to Y_3$ form a degraded broadcast channel. We establish inner and outer
bounds for this case and show that they are tight when the channel $X \to Y_1$
is {\em less noisy} than the channel $X \to Y_3$, which is a more general class
than degraded broadcast channels~\cite{kom75}. We then extend our results to the
case of 3 degraded message sets, where $M_0$ is to be sent to all receivers,
$M_1$ is to be sent to receivers $Y_1$ and $Y_2$ and  $M_2$ is to be sent only
to receiver $Y_1$. A special case of this inner bound gives an inner bound to
the capacity of the 2 degraded message set scenario where $M_0$ is to be sent to
all receivers and $M_1$ is to be sent to receivers $Y_1$ and $Y_2$ only.

\subsection{Inner and Outer Bounds for 2 Degraded Message Sets}
We use superposition coding, indirect decoding, and the Marton achievability
scheme for the general 2-receiver broadcast channels~\cite{mar79} to establish
the following inner bound.
\begin{proposition}
\label{prop:innerbound}
A rate pair $(R_1,R_2)$ is achievable in a general 3-receiver broadcast channel
with 2 degraded message sets if it satisfies the following inequalities: 
\begin{align}
\label{eq:inner}
R_0 &\leq \min\{I(U_2;Y_2), I(U_3;Y_3)\}, \nonumber\\
2R_0 &\leq I(U_2;Y_2) + I(U_3;Y_3) - I(U_2;U_3|U_1), \nonumber\\
R_1 &\leq \min\{ I(X;Y_1 | U_2) + I(X;Y_1 | U_3), I(X;Y_1|U_1)\}, \\ \nonumber
R_0 + R_1 & \leq \min\{ I(X;Y_1),I(U_2;Y_2) + I(X;Y_1|U_2),I(U_3;Y_3) +
I(X;Y_1|U_3)\}, \\ \nonumber
2R_0 + R_1 & \leq I(U_2;Y_2) + I(U_3;Y_3) + I(X;Y_1|U_2,U_3) -
I(U_2;U_3|U_1), \\ \nonumber
2R_0 + 2R_1 &\leq I(U_2;Y_2) + I(X;Y_1|U_2) + I(U_3;Y_3) +
I(X;Y_1|U_3) - I(U_2;U_3|U_1),
\end{align}
for some
$p(u_1,u_2,u_3,x)=p(u_1)p(u_2|u_1)p(x,u_3|u_2)=p(u_1)p(u_3|u_1)p(x,u_2|u_3)$ (or
in other words, both $U_1 \to U_2 \to (U_3,X)$ and $U_1 \to U_3 \to (U_2,X)$
form Markov chains). 
\end{proposition}
\medskip

\begin{proof}
The general idea of the proof is to represent $M_0$ by $U_1$, superimpose two
independent
pieces of  information about $M_1$ to obtain $U_2$ and $U_3$, 
respectively, and then superimpose the remaining information about $M_1$ to
obtain $X$. Receiver $Y_1$ decodes $U_1,U_2,U_3,X$, receivers $Y_2$ and $Y_3$ 
find $M_0$ via indirect decoding of $U_2$ and $U_3$, respectively, as in
Theorem~\ref{th:cap}. 
We now provide an outline of the proof 
\medskip

\noindent{\em Code Generation:} Let $R_1=S_{1}+S_{2}+S_{3}$, where the $S_i \ge
0$, $i=1,2,3$ and $T_{2} \ge S_{2}$, $T_{3} \ge S_{3}$.   Fix a probability mass
function of the required form,
$p(u_1,u_2,u_3,x)=p(u_1)p(u_2|u_1)p(x,u_3|u_2)=p(u_1)p(u_3|u_1)p(x,u_2|u_3)$.

Generate $2^{nR_0}$ sequences $U_1^n(m_0)$, $m_0 \in [1,2^{nR_0}]$ distributed
uniformly at random over the set of typical $U_1^n$ sequences. For each
$U_1^n(m_0)$ generate $2^{nT_{2}}$ sequences $U_2^n(m_0,t_2)$, $t_2 \in
[1,2^{nT_2}]$ distributed uniformly at random from the set of conditionally
typical $U_2^n$ sequences, and $2^{nT_{3}}$ sequences $U_3^n(m_0,t_3)$, $t_3
\in [1,2^{nT_3}]$ distributed uniformly at random over the set of conditionally
typical $U_3^n$ sequences. Randomly partition the $2^{nT_{2}}$ sequences
$U_2^n(m_0,t_2)$ into $2^{nS_{2}}$ equal size bins  and the $2^{nT_{3}}$
$U_3^n(m_0,t_3)$ sequences into $2^{nS_{3}}$ equal size bins. To ensure that
each product bin contains a jointly typical pair $(U_2^n(m_0,t_2),
U_3^n(m_0,t_3))$ with arbitrarily high probability, we
require that (see ~\cite{ele81} for the proof)
\begin{equation}
\label{eq:1}
S_{2} + S_{3} < T_{2} + T_{3} - I(U_2;U_3|U_1).
\end{equation}
Finally for each chosen jointly typical pair $(U_2^n(m_0,t_2), U_3^n(m_0,t_3))$
in each product bin $(s_2,s_3)$, generate $2^{nS_{1}}$ sequences of
codewords $X^n(m_0,s_2,s_3,s_1)$, $s_1 \in [1,2^{nS_1}]$ distributed uniformly
at random over the set of conditionally typical $X^n$ sequences.  

\medskip
\noindent{\em Encoding:}

To send the message pair $(m_0,m_1)$, we express $m_1$ by the triple
$(s_1,s_2,s_3)$ and send the codeword $X^n(m_0,s_2,s_3,s_1)$.
\medskip

\noindent{\em Decoding:}
\begin{enumerate}
\item Receiver $Y_1$ declares that $(m_0,s_2,s_3,s_1)$ is sent if it is the
unique rate tuple such that $Y_1^n$ is jointly typical with $((U_1^n(m_0),
U_2^n(m_0,t_2), U_3^n(m_0,t_3), X^n(m_0,s_2,s_3,s_1))$, and $s_2$ is the
product bin number of $U_2^n(m_0,t_2)$ and $s_3$  is the product bin number of
$U_2^n(m_0,t_3)$. Assuming $(m_0,s_1,s_2,s_3)=(1,1,1,1)$ is sent, we partition
the
error event into the following events.
\begin{enumerate}
\item Error event corresponding to $m_0\neq 1$ occurs with arbitrarily small
probability provided
\begin{equation}
\label{eq:6}
R_0 + S_{1} + S_{2} + S_{3} < I(X;Y_1).
\end{equation}
\item Error event corresponding to $m_0=1, s_2 \neq 1, s_3 \neq 1$ occurs with
arbitrarily small probability provided
\begin{equation}
\label{eq:2}
S_1 +S_2+S_3 < I(X;Y_1|U_1).
\end{equation}
\item Error event corresponding to $m_0=1, s_2 = 1, s_3\neq 1$ occurs with
arbitrarily small probability provided
\begin{equation}
\label{eq:4}
S_{1} + S_{3} < I(X;Y_1|U_1,U_2) = I(X;Y_1|U_2).
\end{equation}
The equality follows from the fact that $U_1 \to U_2 \to (U_3,X)$ form a Markov
Chain.
\item Error event corresponding to $m_0=1, s_2 \neq 1, s_3=1$ occurs with
arbitrarily small probability provided
\begin{equation}
\label{eq:3}
S_{1} + S_{2} < I(X;Y_1|U_1,U_3)=I(X;Y_1|U_3).
\end{equation}
The above equality uses the fact that $U_1 \to U_3 \to (U_2,X)$ forms a Markov
chain.
\item Error event corresponding to $m_0=1, s_2 = 1, s_3=1, s_1 \neq 1$ occurs
with arbitrarily small probability provided
\begin{equation}
\label{eq:5}
S_{1}  < I(X;Y_1|U_1,U_2,U_3)= I(X;Y_1|U_2,U_3).
\end{equation}
Note that the equality here uses a weaker Markov structure $U_1 \to (U_2,U_3)
\to X$.
\end{enumerate} 
\medskip
Thus receiver $Y_1$ decodes $(m_0,s_2,s_3,s_1)$ with arbitrarily small
probability of error
provided equations \eqref{eq:6}-\eqref{eq:5} hold.
\item Receiver $Y_2$ decodes $m_0$ via list decoding of $U_2^n(m_0,t_2)$ (as in
Theorem~\ref{th:cap}). This can be achieved with arbitrarily small probability
of error provided
\begin{equation}
\label{eq:7}
R_0 + T_{2} < I(U_2;Y_2).
\end{equation}
\item Receiver $Y_3$ decodes $m_0$ via list decoding of $U_2^n(m_0,t_3)$ (as in
Theorem~\ref{th:cap}). This can be achieved with arbitrarily small probability
of error provided
\begin{equation}
\label{eq:8}
R_0 + T_{3} < I(U_3;Y_3).
\end{equation}
\end{enumerate}
Combining equations \eqref{eq:1}-\eqref{eq:8} and using the Fourier-Motzkin
procedure~\cite{sch86} to eliminate $T_2,T_3,S_1,S_2$, and $S_3$, we obtain the 
inequalities in \eqref{eq:inner}. The details are given in Appendix
\ref{se:append}.
\end{proof}

\medskip

\noindent{\em Remarks:} 
\begin{enumerate}
\item The above achievability scheme  can be adapted to any joint distribution
$p(u_1,u_2,u_3,x)$. However by letting $\tilde U_2 = (U_2,U_1)$ and letting
$\tilde U_3=(U_3,U_1)$ we observe that the region remains unchanged. Hence,
without
loss of generality we assume the structure of the auxiliary random variables as
described in the proposition. It is also interesting to note that the auxiliary
random variables in the outer bound described in the next subsection also
possess the same structure.

\item An interesting choice of the auxiliary random variables is to set $U_2$ or
$U_3$ equal to $U_1$ (i.e., only one of the the receivers tries to indirectly
decode $M_0$), say let $U_3=U_1$. This reduces the inequalities
~\ref{prop:innerbound} 
(after removing the redundant ones) to: 
\begin{align}
R_0 & \leq \min\{I(U_2;Y_2), I(U_1;Y_3)\}, \nonumber\\
R_1 & \leq I(X;Y_1|U_1), \label{eq:setu3u1}\\
R_0 + R_1 & \leq \min \{I(X;Y_1),I(U_2;Y_2) + I(X;Y_1|U_2),I(U_1;Y_3) +
I(X;Y_1|U_1)\},\nonumber
\end{align}
where $U_1 \to U_2 \to X$ form a Markov chain.

This region includes the capacity region of the multilevel case in
Theorem~\ref{th:cap}. It is easy to verify that for any $U_1\to U_2 \to
X$ that form a Markov chain, the corner points of the region in
Theorem~\ref{th:cap} satisfy the above
inequalities (and this suffices by convexity). 
\end{enumerate}

\medskip

We now establish the following outer bound
\begin{proposition}
\label{prop:outerbound}
Any achievable rate pair $(R_0,R_1)$ for the general 3-receiver broadcast
channel with 2 degraded message sets must satisfy the conditions:
\begin{align*}
R_0 &\leq \min \{I(U_1;Y_1), I(U_2;Y_2) - I(U_2;Y_1|U_1), I(U_3;Y_3) -
I(U_3;Y_1|U_1)\}, \\
R_1 &\leq I(X;Y_1|U_1).
\end{align*}
for some
$p(u_1,u_2,u_3,x)=p(u_1)p(u_2|u_1)p(x,u_3|u_2)=p(u_1)p(u_3|u_1)p(x,u_2|u_3)$,
i.e., the same structure of the auxiliary random variables as in Lemma
\ref{prop:innerbound}. Further one can restrict the cardinalities of
$U_1,U_2,U_3$
to: $\| \mathcal{U}_1 \| \leq \| \mathcal{X} \| + 6$, $ \| \mathcal{U}_2 \|
\leq (\| \mathcal{X} \| + 1)(\| \mathcal{X} \| + 6)$, and $ \| \mathcal{U}_3
\| \leq (\| \mathcal{X} \| + 1)(\| \mathcal{X} \| + 6)$.
\end{proposition}

\medskip
\begin{proof}
The proof follows largely standard arguments. The auxiliary random variables are
identified as $U_{1i} = (M_0,Y_1^{i-1})$, $ U_{2i}=(U_{1i},Y_{2~i+1}^{~n})$,
$U_{3i}=(U_{1i},Y_{3~i+1}^{~n})$. With this identification inequalities $R_0
\leq I(U_1;Y_1)$ and $R_1 \leq I(X;Y_1|U_1)$ is immediate. The other two
inequalities also follow from standard arguments and is briefly outlined here.
\begin{align*}
nR_0 &\leq n \lambda_n + \sum_i I(M_0;Y_{3i}|Y_{3~i+1}^{~n}) \\
& \leq n\lambda_n + \sum_i I(M_0,Y_{3~i+1}^{~n},Y_1^{i-1};Y_{3i}) -
I(Y_1^{i-1};Y_{3i}|M_0, Y_{3~i+1}^{~n}) \\
& \stackrel{(a)}{=} n\lambda_n + \sum_i I(M_0,Y_{3~i+1}^{~n},Y_1^{i-1};Y_{3i}) -
I(Y_{3~i+1}^{~n} ;Y_{1i}|M_0,Y_1^{i-1}) \\
& = n\lambda_n + \sum_i I(U_{3i};Y_{3i}) - I(U_{3i};Y_{1i}|U_{1i}),
\end{align*}
where $(a)$ uses the Csisz\'{a}r sum equality.

The cardinality bounds are established using a similar argument as in
\ref{sse:card}. To create a set of new auxiliary random variables with the
bounds of Proposition \ref{prop:outerbound}, we first replace $U_2$ by
$(U_2,U_1)$ and $U_3$ by $(U_3,U_1)$. It is easy to see from the Markov chain
relationships $U_1 \to U_2 \to (U_3,X)$ and  $U_1 \to U_3 \to (U_2,X)$ that the
following region is same as the that of Proposition \ref{prop:outerbound}.
\begin{align}
R_0 &\leq \min\{ I(U_1;Y_1), I(U_1,U_2;Y_2) + I(X;Y_1|U_1,U_2) - I(X:Y_1|U_1),
\nonumber \\
&\qquad I(U_1,U_3;Y_3) + I(X;Y_1|U_1,U_3) - I(X:Y_1|U_1)\},
\label{eq:eqreg}\\
R_1 &\leq I(X;Y_1|U_1). \nonumber 
\end{align}
Then using standard arguments one can replace $U_1$ by
$U_1^*$ satisfying $\| \mathcal{U}_1^* \| \leq \| \mathcal{X} \| +
6$, such that the distribution of $X$ and $H(Y_1|U_1)$, $ H(Y_1|U_1,U_2)$, $
H(Y_1|U_ 1,U_3)$, $H(Y_2|U_1)$, $H(Y_2|U_1,U_2)$, $ H(Y_3|U_1)$, and $
H(Y_3|U_1,U_3)$ are preserved. Now for each $U_1^*=u_1$ one can find
$U_2^*(u_1)$ with cardinality less than $\| \mathcal{X} \| + 1$ each such that
the distribution of $X$ conditioned on $U_1^*=u_1$, $H(Y_1|U_1^*=u_1,U_2)$, and
$
H(Y_2|U_1^*=u_1,U_2)$ are preserved. Similarly one can find for each 
$U_1^*=u_1$, a random variable $U_3^*(u_1)$ with cardinality less than
$\| \mathcal{X} \| + 1$ each such that the distribution of $X$
conditioned on $U_1^*=u_1$,  $H(Y_1|U_1^*=u_1,U_3)$, and $ H(Y_3|U_1^*=u_1,U_3)$
are preserved. This yields random variables $U_1^*, U_2^*, U_3^*$ 
that preserve the region in \eqref{eq:eqreg}. (Note that as the distribution
of $X$ conditioned on $U_1 = u_1$ is preserved by both $U_2^*(u_1)$ and
$U_3^*(u_1)$, it is possible to get a consistent triple of random variables
$U_1^*, U_2^*, U_3^*$.)
Finally setting $\tilde{U}_1 = U_1^*$, $\tilde{U}_2=(U_1^*,U_2^*)$ and
$\tilde{U}_3=(U_1^*,U_3^*)$ gives the desired bounds on cardinality as well as
the desired Markov relations.
\end{proof}
\medskip

\noindent{\em Remarks:}
\begin{enumerate}
\item The above outer bound  appears to be very different from the inner bound
of
Proposition~\ref{prop:innerbound}. However, by taking appropriate sums of the
inequalities defining the region of Proposition~\ref{prop:outerbound}, we arrive
at the conditions 
\begin{align*}
R_0 & \leq \min\{ I(U_2;Y_2) - I(U_2;Y_1|U_1),I(U_3;Y_3) - I(U_3;Y_1|U_1) ),\\
R_1 & \leq I(X;Y_1|U_1), \\
R_0 + R_1 &\leq \min\{I(X;Y_1), I(U_2;Y_2) + I(X;Y_1|U_2), I(U_3;Y_3) +
I(X;Y_1|U_3)\},\\ 
2R_0 + R_1 &\leq I(U_2;Y_2) + I(U_3;Y_3) + I(X;Y_1|U_2,U_3) - I(U_2;U_3|U_1) +
I(U_2;U_3|Y_1,U_1).
\end{align*}
These conditions include some redundant ones, but are closer in structure to the
inequalities defining the inner bound of
Proposition~\ref{prop:innerbound}.

\item The outer bound in Proposition \ref{prop:outerbound} reduces to the
capacity region for the multilevel case in Theorem~\ref{th:cap}. To see this
observe that when $X \to Y_1 \to
Y_3$ form a Markov chain, 
\begin{equation}
\label{eq:lessnoisyone}
R_0 \leq I(U_3;Y_3) - I(U_3;Y_1|U_1) \leq I(U_3;Y_3) - I(U_3;Y_3|U_1) =
I(U_1;Y_3).
\end{equation}

Further from  $R_1 \leq I(X;Y_1|U_1)$,  we have $R_0 + R_1
\leq I(U_1;Y_3) + I(X;Y_1|U_1)$. Thus the outer bound is contained
in the achievable region of Theorem \ref{th:cap}, i.e.,
\begin{align}
\label{eq:outmod}
R_0 &\leq \min \{I(U_1;Y_3), I(U_2;Y_2)\},\\
R_0 + R_1 & \leq \{I(U_1;Y_3) + I(X;Y_1|U_1),I(U_2;Y_2) +
I(X;Y_1|U_2)\}.\nonumber
\end{align}
\item The inner and outer bounds match if $Y_1$ is less noisy than
$Y_3$~\cite{kom75}, that is if $I(U;Y_3) \le I(U;Y_1)$ for all $p(u)p(x|u)$. As
shown in ~\cite{kom75}, this condition is more general than degradedness. As
such, it defines a larger class than multilevel  broadcast channels.
\medskip
\begin{proposition}
\label{prop:caplessnoisy}
The capacity region for the 3-receiver broadcast channel with 2 degraded message
sets when
$Y_1$ is a {\em less noisy} receiver than $Y_3$ is given by the set
of rate pairs $(R_0,R_1)$  such that
\begin{align}
\label{eq:capln}
R_0 &\leq \min \{I(U_1;Y_3), I(U_2;Y_2)\},\\
R_0 + R_1 &\leq \min \{I(U_1;Y_3) + I(X;Y_1|U_1), I(U_2;Y_2) + I(X;Y_1|U_2)\},
\nonumber
\end{align}
for some $p(u_1)p(u_2|u_1)p(x|u_2)$. 
\end{proposition}

\medskip

From the definition of less noisy receivers \cite{kom75} we have 
$I(U_3;Y_3|U_1=u_1) \le I(U_3;Y_1|U_1=u_1)$ for every choice
of $u_1$ and thus $I(U_3;Y_3 |U_1) 
\leq I(U_3;Y_1|U_1)$ for every $p(u_1)p(u_3|u_1)p(x|u_3)$.
Using \eqref{eq:lessnoisyone} it follows that the general outer bound is
contained in \eqref{eq:capln}.
The corner point of \eqref{eq:capln} (under the less noisy assumption) is
contained in the region given by
\eqref{eq:setu3u1} and thus achievable by setting $U_3=U_1$ in the region of 
Proposition
\ref{prop:innerbound}.
\end{enumerate}

\subsection{Inner Bound for 3 Degraded Message Sets}
\label{sub:3}
In this section we establish an inner bound to the capacity region of the
broadcast channel with 3
degraded message sets where $M_0$ is to be sent to all three receivers, $M_1$ is
to be sent only to $Y_1$ and $Y_2$, and $M_2$ is to be sent only to $Y_1$. We
then specialize the result to the case of 2 degraded message sets scenario,
where $M_0$ is to be sent to all receivers and $M_1$ is to be sent to $Y_1$ and
$Y_2$ and establish optimality for two classes of channels.

The inner bound we establish is closely related to that of
Proposition \ref{prop:innerbound}. To explain the connection, consider
a 3-receiver broadcast channel scenario where message $M_{0}$ is to be sent to
all three
receivers,  message
$M_{12}$ is to be sent to receivers $Y_{1}$ and $Y_{2}$, message $M_{13}$ is to
be sent to receivers $Y_{1}$ and $Y_{3}$, and message $M_{11}$ is to be sent
only to receiver $Y_1$. An inner bound to the capacity region for this scenario
that uses superposition coding and Marton's coding scheme would be to represent
$M_0$ by an auxiliary random variable $U_1$, $(M_0,M_{12})$ by an auxiliary
random variable $U_2$, $(M_0,M_{13})$ by $U_3$, and $(M_0,M_{12}.M_{13},M_{11})$
by $X$, where $U_1 \to U_2 \to (U_3,X)$ and $U_1 \to U_3 \to (U_2,X)$  form
Markov chains. 

The inner bound of Proposition~\ref{prop:innerbound}  follows from
the above scenario by relaxing the conditions that $Y_2$ needs to decode
$M_{12}$ and
$Y_3$ needs to decode $M_{13}$ and considering both messages as parts of the
private message to receiver $Y_1$. However, instead of eliminating the auxiliary
random variables $U_2$ and $U_3$ completely (as in the BZT region, which is a
straightforward extension of the K\"{o}rner-Marton scheme), we keep them and
have
receivers $Y_2$ and $Y_3$ use the new technique of indirect decoding to find
$M_0$ through $U_2$ and $U_3$, respectively. As we have shown in Section
\ref{se:bec}, having these random variables $U_2$
and $U_3$ can strictly improve the achievability region of the 2-message sets
scenario. 

\medskip

Now consider the 3 degraded message set scenario. We relax the condition in the
above scenario  that $Y_3$ needs to decode $M_{13}$. Recall the proof of
Proposition \ref{prop:innerbound}. We let $R_1 = S_2$, $R_2=S_3+S_1$ and
represent $M_0$ by $U_1$, $(M_0,M_1)$ by $U_2$, $(M_0,S_3)$ by $U_3$, and
$(M_0,M_1,M_2)$ by $X$. Receiver $Y_1$ finds $(M_0,M_1,M_2)$ by decoding
$U_1,U_2,U_3,X$; receiver $Y_2$ finds $(M_0,M_1)$ by decoding $U_1,U_2$; and
receiver $Y_3$ finds $M_0$ by indirectly decoding $U_1$ through $U_3$. 
We obtain the following conditions for achievability of any rate tuple
$(R_0,R_1,S_3,S_1)$ by replacing $S_2$ by $R_1$ in conditions
~\eqref{eq:1}-\eqref{eq:8}  and adding the condition $T_2 < I(U_2;Y_2|U_1)$ (to
enable $Y_2$ to completely decode $U_2$).  
\begin{align}
R_1 &\leq T_2,\nonumber\\
S_3 &\leq T_3, \nonumber\\
R_1 + S_{3} &\leq T_{2} + T_{3} - I(U_2;U_3|U_1), \nonumber\\
R_0 + S_{1} + R_1 + S_{3} &\leq I(X;Y_1), \nonumber \\
S_{1} + S_{3} &\leq I(X;Y_1|U_1,U_2) = I(X;Y_1|U_2), \nonumber\\
S_{1} + R_1 &\leq I(X;Y_1|U_1,U_3)=I(X;Y_1|U_3), \label{eq:condgeninner}\\
S_1 + R_1 + S_3 &\leq I(X;Y_1|U_1), \nonumber\\
S_{1}  &\leq I(X;Y_1|U_1,U_2,U_3)= I(X;Y_1|U_2,U_3), \nonumber\\
R_0 + T_{2} &\leq I(U_1,U_2;Y_2) = I(U_2;Y_2), \nonumber\\
T_2 &\leq I(U_2;Y_2|U_1), \nonumber\\
R_0 + T_{3} &\leq I(U_1,U_3;Y_3) = I(U_3;Y_3), \nonumber
\end{align}
for some
$p(u_1,u_2,u_3,x)=p(u_1)p(u_2|u_1)p(x,u_3|u_2)=p(u_1)p(u_3|u_1)p(x,u_2|u_3)$.

Performing Fourier-Motzkin procedure to eliminate the variables $S_1,S_3,T_2$
and $T_3$ yields the following achievable region. 

\begin{theorem}
\label{th:geninner}
A rate triple $(R_0,R_1,R_2)$ is achievable in a general 3-receiver broadcast
channel
with 3 degraded message sets if it satisfies the conditions: 
\begin{align}
R_0  &\leq I(U_3;Y_3) \nonumber \\
R_1 &\leq \min\{ I(U_2;Y_2|U_1), I(X;Y_1|U_3)\}, \nonumber  \\
R_2  &\leq I(X;Y_1|U_2)  \nonumber \\
R_0 + R_1 &\leq  \min \{ I(U_2;Y_2),I(U_2;Y_2|U_1) + I(U_3;Y_3)  -
I(U_2;U_3|U_1)\}, \nonumber\\
2R_0 + R_1  &\leq I(U_2;Y_2)  +  I(U_3;Y_3)  - I(U_2;U_3|U_1), \nonumber\\
R_0 + R_2 &\leq  I(U_3;Y_3) + I(X;Y_1|U_2,U_3) \nonumber \\
R_1 + R_2  &\leq I(X;Y_1|U_1), \nonumber \\
R_0 + R_1 + R_2 &\leq \min \{ I(X;Y_1), I(U_3;Y_3) + I(X;Y_1|U_3), \nonumber \\
&\qquad I(U_2;Y_2|U_1) + I(U_3;Y_3) +
I(X;Y_1|U_2,U_3)  -
I(U_2;U_3|U_1)\}, \label{eq:geninner} \\
2R_0 + R_1 + R_2 &\leq I(U_2;Y_2)  +  I(U_3;Y_3) + I(X;Y_1|U_2,U_3) -
I(U_2;U_3|U_1), \nonumber \\
R_0 + 2R_1  + R_2 &\leq I(U_2;Y_2|U_1) + I(U_3;Y_3) + I(X;Y_1|U_3) -
I(U_2;U_3|U_1), \nonumber \\
2R_0 + 2R_1  + R_2 &\leq I(U_2;Y_2)  +  I(U_3;Y_3)  + I(X;Y_1|U_3) -
I(U_2;U_3|U_1). \nonumber 
\end{align}
for some
$p(u_1,u_2,u_3,x)=p(u_1)p(u_2|u_1)p(x,u_3|u_2)=p(u_1)p(u_3|u_1)p(x,u_2|u_3)$
(i.e., as before both $U_1$ $\to U_2$ $\to (U_3,X)$ and $U_1 \to U_3 \to
(U_2,X)$
form Markov chains).

\end{theorem}

\medskip

\noindent{\em Remark:}
The region of Theorem~\ref{th:geninner} reduces to the inner bound of
Proposition~\ref{prop:innerbound} by setting $R_1=0$. The equivalence between
the
two descriptions is proved in Appendix~\ref{ap:equiv}.

We now consider a 2 degraded message set scenario where $M_0$ is to be sent to
all receivers and $M_1$ is to be sent to receivers $Y_1$ and $Y_2$. The
following inner bound follows from  Theorem~\ref{th:geninner} by setting
$R_2=0$.

\begin{corollary}
\label{co:innerde2me}
A rate pair $(R_0,R_1)$ is achievable in a 3-receiver broadcast channel with 2
degraded message sets, where $M_0$ is to be decoded by all three receivers and
$M_1$ is to be decoded only by $Y_1$ and $Y_2$ if it satisfies the following
conditions:
\begin{align}
R_0  &\leq I(U;Y_3), \nonumber\\
R_1 &\leq \min \{I(X;Y_2|U), I(X;Y_1|U)\},  \label{eq:sc1}\\
R_0 + R_1 &\leq  \min \{I(X;Y_2), I(X;Y_1)\}, \nonumber
\end{align}
for some $p(u)p(x|u)$.
\end{corollary}

\medskip
\noindent{\em Remarks:} 
\begin{enumerate}
\item Region \eqref{eq:sc1} coincides with the
straightforward extension of the K\"{o}rner-Marton 2-receiver region. 
\item By setting $R_2=0$, $U_2=X$, and $U_3=U_1=U$ the region in Theorem
\ref{th:geninner}
reduces to \eqref{eq:sc1}. Thus region in \eqref{eq:sc1} is contained in region
\eqref{eq:geninner}.
\item It may seem that the region obtained by setting $R_2=0$ in
\eqref{eq:geninner} is larger than region \eqref{eq:sc1}, but they are in fact
equal. 
variables, we see that  
Therefore, there is no need to introduce $U_3$. 
To prove this, observe that
\begin{align*}
R_0 + R_1 & \leq I(U_2;Y_2|U_1) + I(U_3;Y_3)  - I(U_2;U_3|U_1) \\
& = I(U_3;Y_3) + I(U_3;Y_2|U_1) + I(U_2;Y_2|U_3) - I(U_3;Y_2|U_2) -
I(U_3;U_2|U_1) \\
& = I(U_3;Y_3) + I(U_2;Y_2|U_3) - I(U_3;U_2|Y_2,U_1) \\
& \leq I(U_3;Y_3) + I(X;Y_2|U_3).
\end{align*}
Thus the rate pairs must satisfy the following inequalities
\begin{align}
R_0 &\leq I(U_3;Y_3), \nonumber \\
R_0 + R_1 &\leq \min \{I(U_3;Y_3) + I(X;Y_2|U_3), I(U_3;Y_3) + I(X;Y_1|U_3) \},
\label{eq:redu2deg} \\
R_0 + R_1 & \leq \min \{I(X;Y_2), I(X;Y_1)\}. \nonumber
\end{align}
Clearly this is contained inside region \eqref{eq:sc1} and hence region
\eqref{eq:geninner} reduces to the one in Corollary \ref{co:innerde2me} when
$R_2=0$.
\item Inner bound  \eqref{eq:sc1} is optimal for the following two special
classes of broadcast channels.
\medskip

\begin{proposition}
\label{prop:deterministic}
Achievable region \eqref{eq:sc1} is tight for deterministic 3-receiver broadcast
channels. 
\end{proposition}
\medskip

It is straightforward to show that the set of  rate pairs $(R_0,R_1)$ such that
\begin{align*}
R_0 &\leq \min\{H(Y_1),H(Y_2),H(Y_3)\},\\
R_0 + R_1 &\leq \min\{H(Y_2),H(Y_1)\},
\end{align*}
for some $p(x)$ constitutes an outer bound on the capacity region. To show
achievability, we need only consider the three choices for $U$: (i) $U=Y_3$,
and (ii) $U=X$, and (iii) $U = \emptyset$.

\medskip

\begin{proposition}
\label{prop:lessnoisy2}
Achievable region \eqref{eq:sc1} is optimal when $Y_1$ is a less
noisy receiver than $Y_3$ and $Y_2$ is a less noisy receiver than $Y_3$. 
\end{proposition}
\medskip

Note that this result generalizes Theroem 3.2
in \cite{dit06} where the authors assume the receivers are $Y_2$ and $Y_1$ are
degraded versions of $Y_3$.
To show optimality, we set $U_i=(M_0,Y_3^{i-1})$ and thus the only non-trivial
inequality in the converse
is $R_1 \leq \min \{I(X;Y_1|U), I(X;Y_2|U)\} $. To see this observe that
\begin{align*}
n R_1 & \leq \sum_i I(M_1;Y_{1i}|M_0,Y_{1~i+1}^n) \\
& \leq \sum_i I(M_1;Y_{1i}|M_0,Y_{1~i+1}^n,Y_3^{i-1}) + \sum_i I(Y_3^{i-1};
Y_{1i}|M_0,Y_{1~i+1}^n)) \\
& \stackrel{(a)}{=} \sum_i I(M_1,Y_{1~i+1}^n;Y_{1i}|M_0,Y_3^{i-1}) - \sum_i
I(Y_{1~i+1}^n;Y_{1i}|M_0,Y_3^{i-1}) + \sum_i I(Y_{1~i+1}^n;Y_{3i}|M_0,Y_3^{i-1})
\\
& \stackrel{(b)}{\leq} \sum_i I(X_i;Y_{1i}|M_0,Y_3^{i-1}),
\end{align*}
where $(a)$ uses the Csisz\'{a}r sum equality and $(b)$ uses the assumption that
$Y_1$ is a less noisy than $Y_3$, 
which implies  that
$I(Y_{1~i+1}^n;Y_{3i}|M_0,Y_3^{i-1}) \leq I(Y_{1~i+1}^n;Y_{1i}|M_0,Y_3^{i-1})$.
The bound $R_1 \leq I(X;Y_2|U)$ can be proved similarly.
\end{enumerate}

\subsection{Inner Bounds for $k$-receiver Broadcast Channels}

The inner bounds discussed in previous subsections suggest the following
extension to general $k$-receiver broadcast channel scenarios with given message
requirements. To illustrate our procedure we shall use the running example of a
3-receiver broadcast channel with  3 messages to receiver subsets:  $\{1\}$,
$\{1,2\}$, and $\{2,3\}$.

To obtain an inner bound to capacity for a given message requirement, we first
consider all nonempty receiver subsets. Let $\mathcal{S}_D$ be the collection of
subsets specified by the message requirements. For each $A \in \mathcal{S}_D$, 
we introduce an auxiliary random variable for every $B \supset A$. Thus in our
example, $\mathcal{S}_D = \left\{ \{1\}, \{2,3\},  \{1,2\} \right\}$, and five
auxiliary random variables are introduced corresponding to the subsets:
$\{1,2,3\}$, $\{1,2\}$, $\{1,3\}$, $\{2,3\}$, and $\{1\}$. Let $\mathcal{S}_I$
denote the receiver subsets for which auxiliary random variables are introduced
but are not in $\mathcal{S}_D$. In the example, $\mathcal{S}_I = \left\{
\{1,3\}, \{1,2,3\} \right\}$.

The receiver subsets with auxiliary random variables assigned to them are
classified into  levels based on their cardinality with the lowest level subsets
having the largest cardinality. There is a Markov structure between the
variables as follows: if $U_B$ represents the auxiliary random variable
corresponding to the subset $B$ and $U_A$ represents the auxiliary random
variable corresponding to the subset $A \subset B$, then one can set
$U_A=(U_B,\tilde U_A)$. Thus an auxiliary random variable $U_A$ corresponding to
a subset $A$ should contain {\em all} auxiliary random variables corresponding
to  the subsets $B \supset A$. For the running  example, Level 1 contains the
subset $\{1,2,3\}$, Level 2 contains the subsets $\{1,2\}$, $\{1,3\}$, and
$\{2,3\}$,  and Level 3 contains the subset $\{1\}$. The Markov relationships
between these auxiliary random variables are defined by:
\begin{align*}
U_{12} &=(U_{123}, \tilde U_{12}), ~ U_{13}=(U_{123}, \tilde U_{13}), ~
U_{23}=(U_{123},\tilde U_{23}),\\
U_{1} &= \{U_{12},U_{13}, \tilde U_{1}\}.
\end{align*} 
Code generation proceeds one level at a time beginning with the lowest level
followed by the second lowest level, and so on. The codebooks corresponding to
auxiliary random variables at each level are randomly generated conditioned on
codewords at the lower level according to the Markov structure of the auxiliary
random variables. Random binning is performed at each level to find jointly
typical codewords to represent message products.

Decoding is performed at receiver $i$ as follows: let $T_i$ represent the
collection of receiver subsets that contains $i$ for which auxiliary random
variables are introduced.  A subset $A\in T_i$ is said to be minimal if there is
no $B \in T_i$ such that $B \subset A$. Let $\mathcal{T}^{\rm min}_i$ be the
collection of minimal subsets in $T_i$.
For the example we obtain
\begin{align*}
T_1 &= \left\{\{1,2,3\},\{1,2\},\{1,3\},\{1\}\right\} \text{ and } 
\mathcal{T}^{\rm min}_1 = \{1\},\\
T_2 &= \left\{\{1,2,3\},\{1,2\},\{2,3\}\right\} \text{ and } \mathcal{T}^{\rm
min}_2 = \left\{\{1,2\},\{2,3\}\right\},\\
T_3 &= \left\{1,2,3\},\{1,3\},\{2,3\}\right\} \text{ and }  \mathcal{T}^{\rm
min}_3 = \left\{\{1,3\},\{2,3\}\right\}.
\end{align*}
By the Markov structure defined above, it is clear that all the messages for
receiver $i$ are represented by the auxiliary random variables $\mathcal{U}^{\rm
min}_i$, which correspond to the elements of $\mathcal{T}^{min}_i$.
The auxiliary random variables in $\mathcal{U}^{min}_i \cap \mathcal{S}_D$
represent private messages for receiver $i$, while those in $\mathcal{U}^{min}_i
\cap \mathcal{S}_D^c$ contain only parts of private messages. Receiver $i$ uses
indirect decoding to find the private messages encoded into cloud centers by
using the satellite codewords represented by the auxiliary random variables in
$\mathcal{U}^{min}_i$. 

In our running example, receiver $3$ indirectly decodes $U_{23}$ using the pair
$(U_{13},U_{23})$. That is, the rate constraints are such that receiver $3$ may
not be able to uniquely decode $U_{13}$ but is able to decode the correct
$U_{23}$. However, receivers $Y_2$ and $Y_1$ should be able to correctly decode
$(U_{12},U_{23})$ and $U_1$, respectively, and hence these receivers impose the
usual (direct) decoding constraints on the rates. In general, when
$\mathcal{U}^{min}_i \cap \mathcal{S}_I = \emptyset$, indirect decoding is not
needed as in the Examples below, where as in Proposition \ref{prop:innerbound}
indirect decoding is needed.

The following two examples show that the above procedure yields the best known
inner bounds for special classes of broadcast channels.
\medskip

\noindent{\em Example 1:} 2-receiver broadcast channel where $M_1$ is to be
decoded by receiver $Y_1$ and $M_2$ is to be decoded by $Y_2$. We generate 3
auxiliary random variables corresponding to the three non-empty subsets of
$\{1,2\}$: $W$ for $\{1,2\}$, $U$ for $\{1\}$ and $V$ for $\{2\}$.  Setting
$\tilde U=(U,W)$ and $\tilde V=(V,W)$ represents the Markov structure among the
variables. Observe that the auxiliary random variables are exactly as in
Marton's coding scheme and so is the code generation we outlined earlier.

\medskip

\noindent{\em Example 2:} $k$-receiver broadcast channel with 2 degraded message
sets, where $M_0$ is to be decoded by receivers $\{1,\ldots,k\}$ and $M_1$ is to
be decoded by $\{1,\ldots,k-1\}$. The only
subsets that we would assign auxiliary random variables to here are
$\{1,\ldots,k\}$ and $\{1,\ldots,k-1\}$. We thus introduce the auxiliary random
variable $U_1$ for $\{1,\ldots,k\}$ and $U_2$ for $\{1,\ldots,k-1\}$. The region
is then be given by
\begin{align*}
R_0 & \leq I(U_1;Y_k), \\
R_0 + R_1 &\leq I(U_2;Y_i), ~\text{for}~ i=1,\ldots,k-1, \\
R_1 & \leq I(U_2; Y_i |U_1) ~\text{for}~ i=1,\ldots,k-1,
\end{align*}
where $U_1 \to U_2 \to X \to \{Y_1,\ldots,Y_l\}$ form a Markov chain. Clearly in
this case it is optimal to set $U_2=X$, which reduces the region to the
straightforward extension of the K\"{o}rner-Marton scheme.

\medskip

\noindent{\em Remark:} Our procedure can result in an explosion in the number of
auxiliary random variables introduced even in simple scenarios. However, as we
have shown in Section \ref{se:bec}, indirect decoding may be needed to achieve
the capacity region for some classes of channels. Thus the introduction of such
a large number of auxiliary random variables may indeed be necessary in
general.

\medskip

\section{Conclusion}
Recent results and conjectures on the capacity of $(k>2)$-receiver
broadcast channels with degraded message sets ~\cite{bzt07,dit06,pdt07} have
lent support to the general belief that the straightforward extension of the
K\"{o}rner-Marton region for the 2-receiver case is optimal. This paper shows
that this is not the case. We show that the capacity region of the 3-receiver
broadcast channels with 2 degraded message sets can be strictly larger than the
straightforward extension of the K\"{o}rner-Marton region. The achievability
proof uses the new idea of indirect decoding whereby a receiver decodes a cloud
center indirectly through joint typicality with a satellite codeword. Using this
idea, we devise new inner bounds to the capacity of the general 3-receiver
broadcast channel with 2 and 3 degraded message sets and show optimality in some
cases. The structure of the auxiliary random variables in the inner bounds can
be naturally extended to more than 3 receivers. The bounds also provide some
insight into how the Marton achievable rate region may be extended to more than
2 receivers. 

The results in this paper suggest that the capacity of the $k>2$-receiver
broadcast channels with degraded message sets is as at least as hard to find as
the capacity of the general 2-receiver broadcast channel with common and private
message. However, it would be interesting to explore the optimality of our new
inner bounds for classes where capacity is known for the general 2-receiver
case, such as deterministic and vector Gaussian broadcast channels. It would
also be interesting to investigate applications of indirect decoding to other
problems, for example, $3$-receiver broadcast channels with confidential message
sets~\cite{czk78}.

\section*{Acknoledgement}
The authors wish to thank Young-Han Kim for
valuable suggestions that has improved the presentation of this paper.

\clearpage

\appendices

\section{Proof of Propositions  \ref{prop:char-bzt}, \ref{prop:char-cap},
\ref{prop:eval-bzt}, and \ref{prop:eval-cap}}
\label{se:append-bzt}
To prove Propositions  \ref{prop:char-bzt}, \ref{prop:char-cap}, note that it is
straightforward to show that each simplified characterization is contained in
the original region as the characterizations are obtained by using the channels
independently. So we only prove the other non-trivial direction. 
\medskip

\noindent{\em Proof of Proposition  \ref{prop:char-bzt}:}

We prove that for the product broadcast channel given by \eqref{eq:channel} 
the BZT region \eqref{eq:3-BZT} reduces to the expression \eqref{eq:PBZT}.

Consider the first term \eqref{eq:bzt-term1} in the BZT region
\begin{align*}
R_0 \le I(U;Y_3) &= I(U; Y_{31},Y_{32})\\
&=  I(U; Y_{31}) + I(U; Y_{32}|Y_{31})\\
&\le  I(U; Y_{31}) + I(U, Y_{31}; Y_{32})\\
&\le   I(U; Y_{31}) + I(U, Y_{11}; Y_{32}).
\end{align*}
Now set $V_1=U$ and $V_2=(U,Y_{11})$. Thus the above inequality becomes
$$ R_0 \le I(V_1;Y_{31}) + I(V_2;Y_{32}). $$

The second term \eqref{eq:bzt-term2} in the BZT region is  simply given by
\[
R_0 \le I(V_1; Y_{21}).
\]
Finally, consider the last term \eqref{eq:bzt-term3}
\begin{align*}
R_1 \le I(X;Y|U) &= I(X_1,X_2;Y_{11},Y_{12}|U) \\
&= H(Y_{11},Y_{12}|U) - H(Y_{11},Y_{12}|X_1,X_2,U) \\
&= H(Y_{11}|U) + H(Y_{12}|U,Y_{11}) -H(Y_{11}|X_1,U) -H(Y_{12}|X_2,U)\\
&= I(X_1;Y_{11}|U) + H(Y_{12}|U,Y_{11}) - H(Y_{12}|X_2,U,Y_{11})\\
&=  I(X_1;Y_{11}|V_1) + I(X_2;Y_{12}|V_2).
\end{align*}
The fact that $p(v_1)p(v_2) p(x_1|v_1) p(x_2|v_2)$ suffices follows from the
structure of the mutual information terms.  

\medskip

\noindent{\em Proof of Proposition \ref{prop:char-cap}:}

We prove that for the product broadcast channel \eqref{eq:channel}
the capacity region given by Theorem \ref{th:cap} reduces to the expression
\eqref{eq:pcapacity}.

Consider the first term \eqref{eq:cap-term1} in the capacity region
\begin{align*}
R_0 \le I(U_1;Y_3) &= I(U_1; Y_{31},Y_{32})\\
&=  I(U_1; Y_{31}) + I(U_1; Y_{32}|Y_{31})\\
&\le  I(U_1; Y_{31}) + I(U_1, Y_{31}; Y_{32})\\
&\le   I(U_1; Y_{31}) + I(U_1, Y_{11}; Y_{32}).
\end{align*}
Now set $V_{11}=U_1$ and $V_{12}=(U_1,Y_{11})$.

\medskip

The second term \eqref{eq:cap-term2} in the capacity region is $R_0 \le
I(U_2;Y_{21})$.
Now set $V_{21}=U_2$ and from $U_1 \to U_2 \to (X_1,X_2)$ we have $V_{11} 
\to V_{21} \to X_1$. Thus the second term can be rewritten as 
$R_0 \le I(V_{21};Y_{21})$

\medskip

Consider the third term \eqref{eq:cap-term3}
\begin{align*}
R_0 + R_1 & \le I(U_1;Y_3) + I(X;Y_1|U_1) \\
&= I(U_1;Y_{31},Y_{32})+ I(X_1,X_2;Y_{11},Y_{12}|U_1) \\
&\le I(U_1; Y_{31}) + I(U_1, Y_{11}; Y_{32}) + H(Y_{11},Y_{12}|U_1) -
H(Y_{11},Y_{12}|X_1,X_2,U_1) \\
&= I(U_1; Y_{31}) + I(U_1, Y_{11}; Y_{32}) + H(Y_{11}|U_1) \\ 
& \qquad \qquad + H(Y_{12}|U_1,Y_{11}) -H(Y_{11}|X_1,U_1)
-H(Y_{12}|X_2,U_1,Y_{11})\\
&= I(V_{11}; Y_{31}) + I(V_{12}; Y_{32}) + I(X_1;Y_{11}|V_{11}) + I(X_2; Y_{12}
|V_{12}).
\end{align*}

Finally consider the last term \eqref{eq:cap-term4}
\begin{align*}
R_0 + R_1 & \le I(U_2;Y_{21}) + I(X;Y_1|U_2) \\ 
&= I(U_2;Y_{21})+ I(X_1,X_2;Y_{11},Y_{12}|U_2) \\
&= I(U_2; Y_{21}) + H(Y_{11},Y_{12}|U_2) - H(Y_{11},Y_{12}|X_1,X_2,U_2) \\
&\le I(U_2;Y_{21}) + H(Y_{11}|U_2) + H(Y_{12}|U_2,Y_{11}) -H(Y_{11}|X_1,U_2)
-H(Y_{12}|X_2,U_2,Y_{11})\\
&= I(V_{21}; Y_{21}) + I(X_1;Y_{11}|V_{21}) + I(X_2; Y_{12} |U_2,Y_{11})\\
&= I(V_{21}; Y_{21}) + I(X_1;Y_{11}|V_{21}) + I(X_2; Y_{12} |U_2,U_1,Y_{11})\\
&\le I(V_{21}; Y_{21}) + I(X_1;Y_{11}|V_{21}) + I(U_2,X_2; Y_{12} |U_1,Y_{11})\\
&= I(V_{21}; Y_{21}) + I(X_1;Y_{11}|V_{21}) + I(X_2; Y_{12} |V_{12}).
\end{align*}
The fact that $p(v_{11})p(v_{21}) p(x_1|v_{21}) p(v_{12}) p(x_2|v_{12})$ 
suffices follows from the structure of the mutual information terms.

\medskip
\medskip

In the proof of propositions \ref{prop:eval-bzt} and \ref{prop:eval-cap}
we shall make use of the following simple fact about the
entropy function~\cite{cot91}.
\[ 
H(ap, 1-p, (1-a)p) = H(p,1-p) + pH(a,1-a).
\]

\noindent{\em Proof of Proposition~\ref{prop:eval-bzt}:}

We prove that the region given by \eqref{eq:PBZT} reduces to
\eqref{eq:peval-bzt}
for the binary erasure channel described by the example in Section
\ref{sse:example}.

Let $\Prob\{V_1 = i\} = \alpha_i,\ \Prob\{X_1 = 0|V_1=i\} = \mu_i$. Then, 
\begin{align*}
I(V_1;Y_{31}) &= H\left(\sum_i \frac{\alpha_i \mu_i}{6}, \frac 56 , \sum_i
\frac{\alpha_i (1- \mu_i)}{6}\right) - \sum_i \alpha_i H\left(\frac{\mu_i}{6},
\frac
56 , \frac{1-\mu_i}{6}\right) \\
&= \frac 16 H\left(\sum_i \alpha_i \mu_i, \sum_i \alpha_i (1-\mu_i)\right) -
\frac
16 \sum_i \alpha_i H(\mu_i, 1-\mu_i), \\
I(V_1;Y_{21}) &= H\left(\sum_i \alpha_i \mu_i, \sum_i \alpha_i (1-\mu_i)\right)
- \sum_i
\alpha_i H(\mu_i, 1-\mu_i), \\
I(X_1;Y_{11}|V_1) &= \sum_i \alpha_i H\left(\frac{\mu_i}{2}, \frac 12 ,
\frac{1-\mu_i}{2}\right) - \sum_i \alpha_i \mu_i H\left(\frac 12, \frac
12\right) - \sum_i
\alpha_i (1 - \mu_i) H\left(\frac 12, \frac 12\right) \\
&= \frac 12 \sum_i \alpha_i H(\mu_i, 1-\mu_i).
\end{align*}

Similarly, let $\Prob\{V_2 = i\} = \beta_i,\ \Prob\{X_2 = 0|V_2=i\} = \nu_i$. 
Then 
\begin{align*}
I(V_2;Y_{31}) &= \frac 12 H\left(\sum_i \beta_i \nu_i, \sum_i \beta_i
(1-\nu_i)\right) - \frac 12 \sum_i \beta_i H(\nu_i, 1-\nu_i), \\
I(X_2;Y_{12}|V_2) & = \sum_i \beta_i H(\nu_i, 1-\nu_i).
\end{align*}

Now setting $\sum_i \beta_i H(\nu_i, 1-\nu_i) = 1 - q,$ and $ \sum_i \alpha_i
H(\mu_i, 1-\mu_i) = 1-p$, we obtain
\begin{align*}
I(U_1;Y_{31}) & = \frac 16 H\left(\sum_i \alpha_i \mu_i, \sum_i \alpha_i
(1-\mu_i)\right) - \frac 16 \sum_i \alpha_i H(\mu_i, 1-\mu_i)\\
& \leq \frac 16 (1 - (1-p)) = \frac p6, \\
I(U_1;Y_{21}) & = H\left(\sum_i \alpha_i \mu_i, \sum_i \alpha_i (1-\mu_i)\right)
-
\sum_i \alpha_i H(\mu_i, 1-\mu_i) \\
& \leq 1-(1-p) = p, \\
I(X_1;Y_{11}|U_1) &= \frac{1-p}{2}, \\
I(U_2;Y_{31}) &= \frac 16 H\left(\sum_i \alpha_i \mu_i, \sum_i \alpha_i
(1-\mu_i)\right) - \frac 16 \sum_i \alpha_i H(\mu_i, 1-\mu_i)\\
& \leq \frac 12 (1-(1-q)) = \frac q2, \\
I(X_2;Y_{12}|U_2) & = 1-q.
\end{align*}

Therefore, any rate pair in the BZT region must satisfy the conditions
\begin{align*}
R_0 & \leq \min \left\{ \frac p6 + \frac q2, p\right\} ,\\
R_1 & \leq \frac{1-p}{2} + 1-q.
\end{align*}
for some $0 \leq p,q \leq 1$.

\begin{figure}[htpb]
\small
\begin{center}
\begin{psfrags}
\psfrag{a}[rb]{$1/2$}
\psfrag{b}[r]{$1/2$}
\psfrag{c}{$V_1$}
\psfrag{d}{$X_1$}
\psfrag{e}[r]{$0$}
\psfrag{f}[r]{$E$}
\psfrag{g}[r]{$1$}
\psfrag{h}[l]{$0$}
\psfrag{i}[l]{$1$}
\psfrag{l}[r]{$0$}
\psfrag{m}[r]{$E$}
\psfrag{n}[r]{$1$}
\psfrag{o}[l]{$0$}
\psfrag{p}[l]{$1$}
\psfrag{u}[rb]{$1/2$}
\psfrag{v}[r]{$1/2$}
\psfrag{x}{$V_2$}
\psfrag{y}{$X_2$}
\psfrag{A}[r]{$p/2$}
\psfrag{B}[r]{$1-p$}
\psfrag{C}[r]{$p/2$}
\psfrag{D}[r]{$q/2$}
\psfrag{E}[r]{$1-q$}
\psfrag{F}[r]{$q/2$}
\epsfig{figure=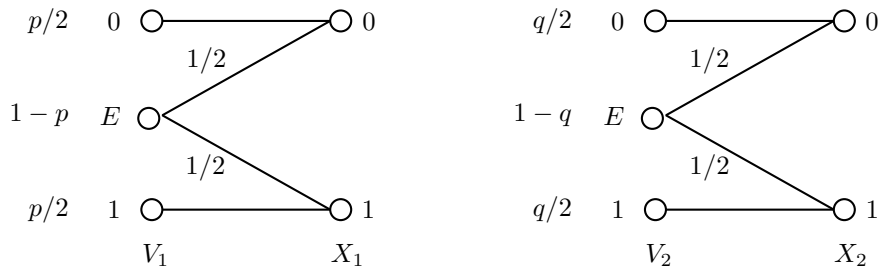,width=0.6\linewidth} \caption{Auxiliary channels that
achieve the boundary of the BZT region.} \label{fig:auxcha}
\end{psfrags}
\end{center}
\end{figure}

It is easy to see that equality is achieved when the marginals of $V_1$ are
given by $\Prob\{V_1 = 0\} = \Prob\{V_1= 1\} = p/2,\ \Prob\{V_1 = E\} =
1-p$ and  the marginals of $V_2$ are given by $\Prob\{V_2 = 0\} = \Prob\{V_2=
1\} = q/2, \Prob\{V_2 = E\} = 1-q$, (see Figure \ref{fig:auxcha}).

\noindent{\em Proof of Proposition \ref{prop:eval-cap}:}
\label{se:append-evalcap}

We prove that the region \eqref{eq:pcapacity} reduces to region
\eqref{eq:peval-cap}
for the binary erasure channel described by the example in Section
\ref{sse:example}.


Assume that
$\Prob\{V_{11}=i\} = \alpha_i, \Prob\{X_1=0|V_{11}=i\} = \mu_i,
\Prob\{V_{12}=i\} = \beta_i, \Prob\{X_2=0|V_{12}=i\}=\nu_i, \Prob\{V_{21}=i\} =
\gamma_i, \Prob\{X_1=0|V_{21}=i\} = \omega_i$.
Further, there exist $r,s,t \in [0,1]$ such that
\begin{align*}
H(X_1|V_{11})&= \sum_i \alpha_i H(\mu_i, 1-\mu_i) = 1 - r,\\ 
H(X_2|V_{12})&= \sum_i \beta_i H(\nu_i, 1-\nu_i) =1-s , \\
H(X_1|V_{21})&= \sum_i \gamma_i H(\omega_i, 1-\omega_i) = 1-t.
\end{align*}
Clearly from the Markov condition $V_{11} \to V_{21} \to X_1$, we require $1-t
\leq 1-r$ or equivalently $r \leq t$.

We can also establish the following in a similar fashion. 
\begin{align*}
I(V_{11}; Y_{31}) &= \frac 16 H\left(\sum_i \alpha_i \mu_i, \sum_i \alpha_i
(1-\mu_i)\right) - \frac 16 \sum_i \alpha_i H(\mu_i, 1-\mu_i) \leq \frac r6, \\
I(V_{12}; Y_{32}) &= \frac 12 H\left(\sum_i \beta_i \nu_i, \sum_i \beta_i
(1-\nu_i)\right) - \frac 12 \sum_i \beta_i H(\nu_i, 1-\nu_i) \leq \frac s2, \\
I(V_{21}; Y_{21}) &=  H\left(\sum_i \gamma_i \omega_i, \sum_i \gamma_i
(1-\omega_i)\right) -  \sum_i \gamma_i H(\omega_i, 1-\omega_i) \leq t, \\
I(X_1;Y_{11}|V_{11}) &= \frac 12 \sum_i \alpha_i H(\mu_i, 1-\mu_i) =
\frac{1-r}{2}, \\
I(X_2; Y_{12} |V_{12}) &= \sum_i \beta_i H(\nu_i, 1-\nu_i) = 1-s, \\
I(X_1;Y_{11}|V_{21}) &= \frac 12 \sum_i \gamma_i H(\omega_i, 1-\omega_i) =
\frac{1-t}{2}.
\end{align*} 

Thus any rate pair in the capacity region must satisfy
\begin{align*}
R_0 & \leq \min \left\{ \frac r6 + \frac s2,  t\right\}, \\
R_0 + R_1 & \leq \min \left\{\frac r6 + \frac s2 + \frac{1-r}{2} + 1-s, t +
\frac{1-t}{2} + 1-s\right\},
\end{align*}
for some $0 \leq r \leq t \leq 1, 0 \leq s \leq 1$.
Note that substituting $r=t$ yields the BZT region.

\begin{figure}[ht]
\small
\begin{center}
\begin{psfrags}
\psfrag{a}[rb]{$1/2$}
\psfrag{b}[r]{$1/2$}
\psfrag{c}{$V_{11}$}
\psfrag{d}{$X_1$}
\psfrag{e}[r]{$0$}
\psfrag{f}[r]{$E$}
\psfrag{g}[r]{$1$}
\psfrag{h}[l]{$0$}
\psfrag{i}[l]{$1$}
\psfrag{l}[r]{$0$}
\psfrag{m}[r]{$E$}
\psfrag{n}[r]{$1$}
\psfrag{o}[l]{$0$}
\psfrag{p}[l]{$1$}
\psfrag{u}[rb]{$1/2$}
\psfrag{v}[r]{$1/2$}
\psfrag{x}{$V_{12}$}
\psfrag{y}{$X_2$}
\psfrag{A}[r]{$r/2$}
\psfrag{B}[r]{$\bar{r}$}
\psfrag{C}[r]{$r/2$}
\psfrag{D}[r]{$s/2$}
\psfrag{E}[r]{$\bar{s}$}
\psfrag{F}[r]{$s/2$}
\psfrag{G}[r]{$(t-r)/2$}
\psfrag{H}[r]{$(t-r)/2$}
\psfrag{I}{$V_{21}$}
\epsfig{figure=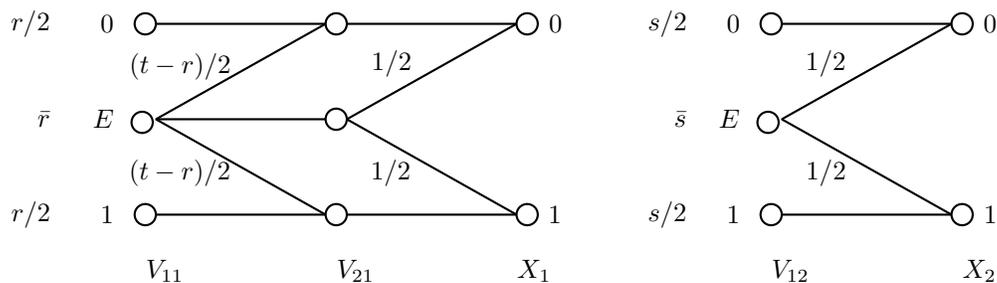,width=0.7\linewidth} \caption{Auxiliary channels
that achieve the boundary of the capacity region.} 
\label{fig:auxcapbec}
\end{psfrags}
\end{center}
\end{figure}

Equality in the above conditions is achieved by the choices of auxiliary random
variables 
shown in Figure \ref{fig:auxcapbec}, and thus the above region is the capacity
region.

\section{Fourier-Motzkin Elimination for Proposition \ref{prop:innerbound}}
\label{se:append}

In this section we provide the details of the Fourier-Motzkin procedure in the
proof of Proposition \ref{prop:innerbound}.

To eliminate $T_2, T_3$ we need to consider the following set of inequalities
\begin{align*}
S_2 & \leq T_2, \\
S_3 & \leq T_3, \\
S_2 + S_3 & \leq T_2 + T_3 - I(U_2; U_3|U_1), \\
R_0 + T_{2} &\leq I(U_2;Y_2), \\
R_0 + T_{3} &\leq I(U_2;Y_3). 
\end{align*}

Elimination $T_2$ first we end up with
\begin{align*}
S_3 & \leq T_3, \\
R_0 + S_2 + S_3 & \leq I(U_2;Y_2) + T_3 - I(U_2; U_3|U_1), \\
R_0 + S_{2} &\leq I(U_2;Y_2), \\
R_0 + T_{3} &\leq I(U_2;Y_3). 
\end{align*}

Elimination of $T_3$ in the above leads us to
\begin{align*}
2 R_0 + S_2 + S_3 & \leq I(U_2;Y_2) + I(U_2;Y_3)   - I(U_2; U_3|U_1), \\
R_0 + S_{2} &\leq I(U_2;Y_2), \\
R_0 + S_{3} &\leq I(U_2;Y_3). 
\end{align*}
Thus any pair $R_0, R_1 = S_1+S_2+S_3$ that satisfies the following set of
inequalities is achievable
\begin{align*}
S_1 &\geq 0, \\
S_2 &\geq 0, \\
S_3 &\geq 0, \\
R_0 + S_{2} &\leq I(U_2;Y_2),  \\
R_0 + S_{3} &\leq I(U_3;Y_3), \\
2R_0 + S_{2} + S_{3} &\leq I(U_2;Y_2) + I(U_3;Y_3) - I(U_2;U_3|U_1), \\
R_0 + S_{2} + S_{3} + S_{1} &\leq I(X;Y_1), \\
S_{2} + S_{3} + S_{1} & \leq I(X;Y_1|U_1), \\
S_{2} + S_{1} & \leq I(X;Y_1|U_3), \\
S_{3} + S_{1} &\leq I(X;Y_1|U_2), \\
S_{1} &\leq I(X;Y_1|U_2,U_3).
\end{align*}
Substituting for $S_1=R_1-S_2-S_3$ yields
\begin{align*}
S_2 &\geq 0, \\
S_3 &\geq 0, \\
S_2 + S_3 &\leq R_1, \\
R_0 + S_{2} &\leq I(U_2;Y_2),  \\
R_0 + S_{3} &\leq I(U_3;Y_3), \\
2R_0 + S_{2} + S_{3} &\leq I(U_2;Y_2) + I(U_3;Y_3) - I(U_2;U_3|U_1), \\
R_0 + R_1 &\leq I(X;Y_1), \\
R_1 & \leq I(X;Y_1|U_1), \\
R_1  & \leq S_3 + I(X;Y_1|U_3), \\
R_1  &\leq S_2 + I(X;Y_1|U_2), \\
R_{1} &\leq S_2 + S_3 + I(X;Y_1|U_2,U_3).
\end{align*}
Elimination of $S_2$ leads to
\begin{align*}
0 &\leq S_3 \\
R_0 + S_{3} &\leq I(U_3;Y_3), \\
R_0 + R_1 &\leq I(X;Y_1), \\
R_1 & \leq I(X;Y_1|U_1), \\
R_1  & \leq S_3 + I(X;Y_1|U_3), \\
S_3 &\leq R_1, \\
R_0  &\leq I(U_2;Y_2),  \\
2R_0 + S_{3} &\leq I(U_2;Y_2) + I(U_3;Y_3) - I(U_2;U_3|U_1), \\
S_3 &\leq  I(X;Y_1|U_2), \\
R_0 + R_1 &\leq I(U_2;Y_2) + I(X;Y_1|U_2), \\
2R_0 + R_1 + S_{3} &\leq I(U_2;Y_2) + I(U_3;Y_3) - I(U_2;U_3|U_1)
+ I(X;Y_1|U_2),\\
0 &\leq  I(X;Y_1|U_2,U_3) ~ \text{redundant,}\\
R_0 + R_{1} &\leq I(U_2;Y_2) + S_3 + I(X;Y_1|U_2,U_3),  \\
2R_0 + R_{1}  &\leq I(U_2;Y_2) + I(U_3;Y_3)  + I(X;Y_1|U_2,U_3) -
I(U_2;U_3|U_1).
\end{align*}
Finally eliminating $S_3$ (and removing redundant inequalities) leads one to
the region in Proposition \ref{prop:innerbound}.

\section{Proof of Remark 1 following Theorem~\ref{th:geninner}}
\label{ap:equiv}
Consider the 3-receiver broadcast channel with 3 degraded message sets. Let $R_2
= S_1+S_2+S_3$. The proof is in three steps:

\begin{itemize}
\item[$(i)$] First, we show that any rate tuple $(R_0,R_1,S_1,S_2,S_3)$ is
achievable provided
\begin{align}
R_1 &\leq T_{21}, \nonumber \\
S_2 & \leq T_{22}, \nonumber \\
S_3 &\leq T_3, \nonumber \\
R_1 + S_3 &\leq T_{21} +  T_{3} - I(\tilde{U}_2;U_3|U_1),\nonumber \\
R_1 + S_2 + S_{3} &\leq T_{21} + T_{22} + T_{3} - I(U_2;U_3|U_1),\nonumber \\
R_0 + S_{1} + R_1 + S_2 + S_{3} &\leq I(X;Y_1),\nonumber  \\
S_1 + R_1 + S_2 + S_3 &\leq I(X;Y_1|U_1),\nonumber   \\
S_{1} + S_2 + S_{3} &\leq I(X;Y_1|\tilde U_2), \label{eq:equiv} \\
S_{1} + S_{3} &\leq I(X;Y_1|U_2), \nonumber \\
S_{1} + R_1 + S_2 &\leq I(X;Y_1|U_3), \nonumber  \\
S_1 + S_2 &\leq I(X;Y_1|U_3,\tilde U_2),\nonumber  \\
S_{1}  &\leq I(X;Y_1|U_2,U_3), \nonumber \\
R_0 + T_{21} + T_{22} &\leq I(U_2;Y_2), \nonumber \\
T_{21} &\leq I(\tilde U_2;Y_2|U_1), \nonumber \\
R_0 + T_{3} &\leq  I(U_3;Y_3), \nonumber
\end{align}
for  $p(u_1, \tilde u_2, u_2, u_3,x)$ $= p(u_1)p(\tilde u_2|u_1)p(u_2|\tilde
u_2)p(x,u_3|u_2) $ $= $ $p(u_1)p(u_3|u_1)p(\tilde u_2,u_2|u_3)p(x|u_2,u_3)$,
i.e. $U_1
\to U_3 \to (\tilde U_2,U_2,X)$ and $U_1 \to \tilde U_2 \to U_2 \to (U_3,X)$
form Markov chains.

\item[$(ii)$] Then, we show that the region defined by \eqref{eq:equiv} is equal
to the
inner bound in Theorem~\ref{th:geninner}. 

\item[$(iii)$] Finally we show that when $R_1=0$, the
conditions \eqref{eq:equiv} reduce to conditions \eqref{eq:1}-\eqref{eq:8}  in
the proof of 
Proposition~\ref{prop:innerbound}, thus completing the proof of Remark 1. 

\end{itemize}

\medskip

\subsection{Achievability of Rates Satisfying \eqref{eq:equiv}}
\label{sse:ap-achi}

First we outline the achievability of any  rate tuple $(R_0,R_1,S_1,S_2,S_3)$
that satisfies conditions \eqref{eq:equiv}. Code generation is very similar to
that in the proof of Proposition~\ref{prop:innerbound}. We insert $\tilde U_2$,
an auxiliary random variable representing the information about $M_1$, between
$U_1$ and $U_2$; so for every $U_1^n(m_0)$ we generate $2^{nT_{21}}$ $
U_2^n(m_0,m_1)$ sequences and randomly partition them into $2^{nR_1}$ bins. For
each $\tilde U_2^n(m_0,m_1)$, we generate $2^{nT_{22}}$ $U_2^n(m_0,m_1,t_{21})$
sequences and randomly partition them into $2^{nS_2}$ bins. We then generate
$2^{nT_{3}}$ $U_3^n(m_0,t_3)$ sequences and partition them into $2^{nS_3}$ bins.
For each product bin $((m_1,s_2),s_3)$ we select a jointly typical pair
$(U_2^n(m_0,m_1,t_{2}), U_3^n(m_0,t_3))$. Finally for product bin
$((m_1,s_2),s_3)$ with corresponding jointly typical $(U_2^n(m_0,m_1,t_{2}),
U_3^n(m_0,t_3))$ pair, we generate $2^{nS_1}$ sequences
$X^n(m_0,m_1,s_{2},s_3,s_1)$.

To ensure correct code generation (existence of relevant jointly typical
sequences) we require that
\begin{align*}
R_1 &\le T_{21},\\
S_2 &\le T_{22}, \\
S_3 &\le T_3, \\
R_1 + S_3 &\leq T_{21} +  T_{3} - I(\tilde{U}_2;U_3|U_1),\nonumber \\
R_1 + S_2 + S_{3} &< T_{21} + T_{22} + T_{3} - I(U_2;U_3|U_1).
\end{align*}
\medskip

Receiver $Y_1$ uses joint typicality to find $(m_0,m_1,s_1,s_2,s_3)$. The
following conditions on the probability of error  ensure successful decoding
(the corresponding events that partition the error event are listed).
\begin{align*}
R_0 + S_{1} + R_1 + S_2 + S_{3} &< I(X;Y_1), & &~(\text{event:}~\hat{m}_0\neq1)
\\
S_1 + R_1 + S_2 + S_3 &< I(X;Y_1|U_1), & &~(\text{event: }~(\hat{m}_0=1,
\hat{m}_1 \neq 1, \hat{s}_3 \neq 1)) \\
S_{1} + S_2 + S_{3} &< I(X;Y_1|\tilde U_2), & &~(\text{event:
}~(\hat{m}_0=1,\hat{m}_1=1,\hat{s}_2\neq(1,1), \hat{s}_3 \neq 1)) \\
S_{1} + S_{3} &< I(X;Y_1|U_2), & &~(\text{event:
}~(\hat{m}_0=1,\hat{m}_1=1,\hat{s}_2=(1,1), \hat{s}_3 \neq 1)) \\
S_{1} + R_1 + S_2 &< I(X;Y_1|U_3), & &~(\text{event:
}~(\hat{m}_0=1,\hat{s}_3=1, \hat{m}_1 \neq 1)) &\\
S_1 + S_2 &< I(X;Y_1|U_3,\tilde U_2), & &~(\text{event:
}~(\hat{m}_0=1,\hat{s}_3=1, \hat{m}_1=1,\hat{s}_2\neq(1,1)))  \\
S_{1}  &< I(X;Y_1|U_2,U_3), & &~(\text{event: }~(\hat{m}_0=1,\hat{s}_3=1,
\hat{m}_1=1, \hat{s}_2=(1,1), \hat{s}_1 \neq 1)).
\end{align*}
Receiver $Y_2$ decodes $m_0$ via indirect decoding using $U_2$ and $m_1$ by
decoding $\tilde U_2$ conditioned on $U_1$. This is successful provided
\begin{align*}
R_0 + T_{21} + T_{22} &< I(U_2;Y_2), \\
T_{21} &< I(\tilde U_2;Y_2|U_1). 
\end{align*}
Receiver $Y_3$ decodes $m_0$ via indirect decoding using $U_3$. This step
succeeds
provided
\begin{align*}
R_0 + T_{3} &<  I(U_3;Y_3).
\end{align*}
Combining the above conditions we see that any rate tuple satisfying
\eqref{eq:equiv} is achievable.

\subsection{Equivalence of Conditions \eqref{eq:equiv} to Theorem
\ref{th:geninner}}

In one direction, setting $\tilde U_2=U_2, S_2=0$, $T_{22}=0$ and $T_{21}=T_2$,
we 
obtain \eqref{eq:condgeninner}. Thus conditions \eqref{eq:equiv} contain the
region 
described by Theorem \ref{th:geninner}. 

For the reverse direction we break down the argument into two cases.
\medskip

{\em Case 1:} $T_{22} < I(U_2;Y_2|\tilde U_2)$ 

Observe that $Y_2$ can also decode $S_2$ and setting $\tilde R_1=R_1+S_2$,
$\tilde R_2=R_2 - S_2$, and $T_{21}+T_{22}=T_2$ we see that conditions
\eqref{eq:equiv} 
along with  $T_{22} < I(U_2;Y_2|\tilde U_2)$ imply conditions
\eqref{eq:condgeninner}.
Thus under $T_{22} < I(U_2;Y_2|\tilde U_2)$, the region described by
\eqref{eq:equiv} is 
contained in the region described by Theorem \ref{th:geninner}. 
\medskip

{\em Case 2:} $T_{22} < I(U_2;Y_2|\tilde U_2)$ 

If $T_{22} \geq I(U_2;Y_2|\tilde U_2)$, then the condition $R_0 + T_{21}+ T_{22}
< I(U_2;Y_2)$
implies that $R_0 + T_{21}  < I(\tilde U_2;Y_2)$ and $Y_2$'s requirement for
successful
decoding can be changed to
\begin{align*}
R_0 + T_{21}  &< I(\tilde U_2;Y_2), \\
T_{21} &< I(\tilde U_2;Y_2|U_1). 
\end{align*}
In the rest of the inequalities, replacing $U_2$ by $\tilde U_2$ only weakens
them and hence it is optimal to set $U_2 = \tilde U_2$. These new inequalities
imply
\eqref{eq:condgeninner} in which we replace $S_1$ by $S_1+S_2$ and $U_2$ by
$\tilde U_2$.
Thus under $T_{22} > I(U_2;Y_2|\tilde U_2)$ also, the region described by
\eqref{eq:equiv} is 
contained in the region described by Theorem \ref{th:geninner}.

Combining Cases 1 and 2 we see that rate pairs satisfying conditions
\eqref{eq:equiv} is contained in the region 
described by Theorem \ref{th:geninner}. This completes the proof of their
equivalence.

\medskip

\subsection{Reduction to Proposition \ref{prop:innerbound}}

If $R_1=0$, the region described by conditions \eqref{eq:equiv} reduce to
\begin{align}
0 &\leq T_{21}, \nonumber \\
S_2 & \leq T_{22}, \nonumber \\
S_3 &\leq T_3, \nonumber \\
S_3 &\leq T_{21} +  T_{3} - I(\tilde{U}_2;U_3|U_1),\nonumber \\
S_2 + S_{3} &\leq T_{21} + T_{22} + T_{3} - I(U_2;U_3|U_1), \nonumber\\
S_{1} + S_2 + S_{3} &\leq  I(X;Y_1|\tilde U_2), \nonumber \\
R_0 + S_{1}  + S_2 + S_{3} &\leq I(X;Y_1), \nonumber \\
S_1 +  S_2 + S_3 &\leq  I(X;Y_1|U_1), \label{eq:reduR10}\\
S_{1} + S_{3} &\leq  I(X;Y_1|U_2), \nonumber \\
S_{1} +  S_2 &\leq  I(X;Y_1|U_3), \nonumber \\
S_1 + S_2 &\leq  I(X;Y_1|U_3,\tilde U_2) \nonumber \\
S_{1}  &\leq I(X;Y_1|U_2,U_3) \nonumber \\
R_0 + T_{21} + T_{22} &\leq   I(U_2;Y_2), \nonumber \\
T_{21} &\leq  I(\tilde U_2;Y_2|U_1) \nonumber\\
R_0 + T_{3} &\leq  I(U_3;Y_3). \nonumber
\end{align}
Recalling that $R_2=S_2+S_3+S_1$, and setting $\tilde
T_{22} = T_{21}+T_{22}$, observe that any $(R_0,S_1,S_2,S_3)$ satisfying the
above inequalities \eqref{eq:reduR10} also satisfies
\begin{align*}
S_2 & \leq \tilde T_{22}, \\
S_3 &\leq T_3, \\
S_2 + S_{3} &\leq \tilde T_{22}  + T_{3} - I(U_2;U_3|U_1),\\
R_0 + S_{1}  + S_2 + S_{3} &\leq I(X;Y_1), \\
S_1 +  S_2 + S_3 &\leq I(X;Y_1|\tilde U_2), \\
S_{1} + S_{3} &\leq I(X;Y_1|U_2), \\
S_{1} +  S_2 &\leq I(X;Y_1|U_3), \\
S_1 + S_2 & \leq I(X;Y_1|U_3,\tilde U_2), \\
S_{1}  &\leq I(X;Y_1|U_2,U_3), \\
R_0 + \tilde T_{22} &\leq  I(U_2;Y_2), \\
R_0 + T_{3} &\leq I(U_3;Y_3).
\end{align*}

These conditions are clearly maximized by setting $\tilde U_2 = U_1$ which in
turn reduces the equations to conditions \eqref{eq:1}-\eqref{eq:8} of
\ref{prop:innerbound}.. Thus the region defined
by \eqref{eq:reduR10} is contained in the region given by Proposition
\ref{prop:innerbound}. The other direction is direct as the region in
Proposition \ref{prop:innerbound}
is obtained by setting $\tilde U_2=U_1$ in \eqref{eq:reduR10}. 
This completes the proof of Remark 1. 

\noindent{\em Remark:} Observe that we do not need the auxiliary random variable
$\tilde U_2$ to characterize the region in either the 3 degraded message sets
case (Theorem \ref{th:geninner}) or the
2 degraded message sets case (Proposition~\ref{prop:innerbound}). This is in
accordance with the structure of auxiliary 
random variables as prescribed by the remark in the introduction of
Subsection~\ref{sub:3}.

\end{document}